\documentclass[12pt]{article}
\usepackage{latexsym,epsf}
\usepackage{epsfig,amsfonts,cite}
\usepackage{amssymb}

\newcommand{\obs}{\mathcal{O}}
\newcommand{\text}{\rm}
\newcommand{\cov}{{\rm Cov}\ }

\begin{document}

\title{Finite-Size Critical Behavior of 
\\ the Driven Lattice Gas
}

\author{
  \\
  { Sergio Caracciolo and Andrea Gambassi }              \\
  {\small\it Scuola Normale Superiore and INFN -- Sezione di Pisa}  \\[-0.2cm]
  {\small\it I-56100 Pisa, ITALIA}          \\[-0.2cm]
  {\small Internet: {\tt sergio.caracciolo@sns.it, andrea.gambassi@sns.it}}   
  \\[-0.1cm]  \and
  { Massimiliano Gubinelli}              \\
  {\small\it Dipartimento di Fisica and INFN -- Sezione di Pisa}    \\[-0.2cm]
  {\small\it Universit\`a degli Studi di Pisa}        \\[-0.2cm]
  {\small\it I-56100 Pisa, ITALIA}          \\[-0.2cm]
  {\small Internet: {\tt mgubi@cibs.sns.it}}   
  \\[-0.1cm]  \and
  { Andrea Pelissetto}              \\
  {\small\it Dipartimento di Fisica and INFN -- Sezione di Roma I}
  \\[-0.2cm] 
  {\small\it Universit\`a degli Studi di Roma ``La Sapienza''}        \\[-0.2cm]
  {\small\it I-00185 Roma, ITALIA}          \\[-0.2cm]
  {\small Internet: {\tt Andrea.Pelissetto@roma1.infn.it}}   \\[-0.2cm]
  {\protect\makebox[5in]{\quad}}  
  \\
}
\vspace{0.5cm}

\newcommand{\be}{\begin{equation}}
\newcommand{\ee}{\end{equation}}
\newcommand{\<}{\langle}
\renewcommand{\>}{\rangle}
\newcommand{\R}[2]{$(#2,#1)$}
\def\spose#1{\hbox to 0pt{#1\hss}}
\def\ltapprox{\mathrel{\spose{\lower 3pt\hbox{$\mathchar"218$}}
 \raise 2.0pt\hbox{$\mathchar"13C$}}}
\def\gtapprox{\mathrel{\spose{\lower 3pt\hbox{$\mathchar"218$}}
 \raise 2.0pt\hbox{$\mathchar"13E$}}}

\newcommand{\co}{ {\cal O}}

\maketitle
\thispagestyle{empty}   

\vspace{0.2cm}

\begin{abstract}
We present a detailed Monte Carlo study of the high-temperature 
phase of the two-dimensional driven lattice gas 
at infinite driving field. We define a finite-volume 
correlation length, study its finite-size-scaling behavior 
and extrapolate it to the infinite-volume limit by using 
an iterative extrapolation method. The same method is applied 
to the susceptibility. We determine the transverse exponents 
$\gamma$ and $\nu$. They turn out to be in perfect agreement
with the theoretical predictions. We also compute the 
transverse Binder parameter, which apparently vanishes at the 
critical point, and the distribution of the 
magnetization. These results confirm the Gaussian nature of the 
transverse excitations.

\end{abstract}

\clearpage

%
%

\newcommand{\reff}[1]{(\ref{#1})}
\newcommand{\kls}{{\sc KLS}}
\newcommand{\su}[2]{\frac{#1}{#2}}
\newcommand{\me}{{\it master equation}}
\newcommand{\db}{{\it detailed balance}}
\newcommand{\ve}[1]{{\bf #1}}

\newcommand{\dlg}{{\sc DLG}}
\newcommand{\fss}{{\sc FSS}}

\newcommand{\pe}[1]{#1_{\bot}}
\newcommand{\pa}[1]{#1_{\|}}
\newcommand{\spint}[1]{\ensuremath{\int {\rm d}^{#1}\ve{x} \, {\rm d}t}}
\newcommand{\imint}[2]{\ensuremath{\int \! \frac{ {\rm
        d}^{#1}\ve{#2}}{(2\pi)^{#1}} \, \frac{{\rm d}\omega}{2\pi}}} 
\newcommand{\nbpa}{\ensuremath{\Delta_{\|}}}
\newcommand{\nbpe}{\ensuremath{\Delta_{\bot}}}
\newcommand{\grpe}{\ensuremath{\vett{\nabla}_{\bot}}}
\newcommand{\grpa}{\ensuremath{\nabla_{\|}}}
\newcommand{\qpeq}{\ensuremath \pe{\ve{q}}^{2}}
\newcommand{\ts}{\ensuremath \tilde{s}}
\newcommand{\G}[2]{\ensuremath \Gamma_{#1\,#2}} 
\newcommand{\tn}{\ensuremath \tilde{n}}

\section{Introduction}

At present, the statistical mechanics of systems in thermal equilibrium is 
quite well established. On the other hand, little is known in general
for non-equilibrium systems. It seems therefore worthwhile to study simple models
which are out of thermal equilibrium. One of such systems is the model introduced
at the beginning of the eighties by 
Katz, Lebowitz, and Spohn~\cite{Katz}, who studied the 
stationary state of a lattice gas under the action of an external
drive. The model, hereafter called {\it driven lattice gas} (\dlg), is
a kinetic Ising model on a periodic domain with Kawasaki dynamics and
biased jump rates. Although not in thermal equilibrium, the \dlg\
has a time-independent stationary state and shows a finite-temperature 
transition, which is however different in nature from its
equilibrium counterpart \cite{foot1}.

Despite its simplicity, the \dlg\  has not been solved exactly \cite{foot1a}. 
Nonetheless, many results have been obtained by means of Monte Carlo
(MC) simulations and by using field-theoretical methods. 
In particular, Refs.~\cite{Janssen86a,Leung86} 
developed a continuum model which is believed to 
capture the basic features of the transition and which provides 
exact predictions for the critical exponents \cite{foot2}.
Several computer 
simulations studied the critical behavior of these systems in two 
and three dimensions~\cite{Zhang88,Leung92,Wang96,Leung99}. These simulations
provided good support to the field-theoretical predictions, once it was
understood that the highly anisotropic character of the transition
required some kind of anisotropic finite-size scaling (FSS). Some
attempts to formulate a phenomenological FSS theory can be found
in Refs.~\cite{Leung92,Binder89}.

In spite of the extensive numerical work, there are no direct studies of the correlation 
length so far, essentially because it is not easy to define it.
Indeed, in the high-temperature phase the model
shows long-range correlations due to the violation of detailed
balance~\cite{Garrido90,Grinstein91}. Therefore, no correlation length can be defined from
the large-distance behavior of the two-point correlation function. A 
different definition is therefore necessary: in Refs. \cite{Valles8687}
a parallel correlation length is defined. However, this definition suffers
from many ambiguities (see the discussion in Ref. \cite{Schmittmann95}) 
and gives results for the exponent $\pa{\nu}$ which are not in agreement 
with the theory \cite{Zhang88}. Even more difficult appears the definition
of a transverse correlation length because of the presence of negative 
correlations at large distances \cite{Zhang88,Schmittmann95}.

In this paper we will define a finite-volume transverse correlation length
generalizing the definition of the second-moment correlation length that is 
used in equilibrium systems. Because of the conserved dynamics, such a 
generalization requires some care. Here, we will use the results of 
Ref.~\cite{correlation_length}.

Once a finite-volume correlation length is defined, we can use the method of 
Ref. \cite{Caracciolo95} to determine infinite-volume quantities \cite{foot3}.
Such a method is particularly efficient and it has been applied successfully 
to many different models 
\cite{Caracciolo95b,Caracciolo:1996ah,Mana-etal-96,Mendes-etal-96,Mana-etal-97,Ferreira-Sokal-99,Palassini-Caracciolo-99}. Once the infinite-volume 
quantities have been computed, we can determine the critical exponents 
by analyzing the singularities in the limit $T \to T_c$.

The paper is organized as follows.
In Sec.~\ref{sec-2} we describe the model and define the
observables measured in the MC simulation.
Section~\ref{sec:fss} reviews the FSS theory and introduces the 
basic formulae that are used in the analysis
of  the simulations.  Next in Sec.~\ref{sec-4}, we consider the 
field-theoretical model of
Refs.~\cite{Janssen86a,Leung86} in a finite 
geometry and compute the
FSS functions for several observables. 
In Sec.~\ref{sec:simulations} we describe the
simulations and present the results which are then discussed in
Sec.~\ref{sec:conclusions}. 
We confirm the field-theoretical predictions with good precision 
both for infinite-volume quantities and for the finite-size behavior 
at the critical
point. Our results also indicate that the 
transverse Binder parameter vanishes at the critical point, in 
agreement with the idea that transverse modes are purely Gaussian.

\section{Definitions}
\label{sec-2}

\subsection{The model}
\label{sec-2.1}
As a prelude to the \dlg\,
consider a simple lattice gas of interacting particles. We
consider a finite square lattice $\Lambda$\ 
and $N$\ particles, each of them occupying a different lattice site. A
configuration of the system is specified by the set of
occupation numbers of each site 
$
        n = \{n_{i} \in \{0,1\}\}_{i\in\Lambda}.
$
We assume a nearest-neighbor attractive (``ferromagnetic'' in spin language)
Hamiltonian 
\begin{equation}
        H_{\Lambda}[n]=-J\sum_{\langle i,j \rangle\in\Lambda} n_{i}\,n_{j},
\end{equation} 
where $J>0$ (in the following we will set without loss of generality 
$J=1$) and the sum runs over all lattice nearest neighbors. 
We consider a discrete-time Kawasaki-type dynamics~\cite{Kawasaki},
which preserves the total number of particles $N$ or, 
equivalently, the density 
\begin{equation}
        \rho_{\Lambda} \equiv  \su{1}{|\Lambda|}\sum_{i\in\Lambda} n_{i},
\end{equation}
where $|\Lambda|$\ is the total number of sites in $\Lambda$.
At each step, we choose randomly a lattice link $\langle i,j \rangle$. 
If $n_i = n_j$, nothing happens. Otherwise, we propose a swap, i.e. 
a particle jump, with probability $w(\Delta H/T)$, where 
\begin{equation}
        \Delta H = H_\Lambda[n'] - H_\Lambda[n]
\end{equation}
is the difference in energy between the new ($n'$) and the old ($n$) 
configuration. If the probability $w(x)$ satisfies
\begin{equation}
        w(-x)=e^{x}\,w(x),
\end{equation} 
then the dynamics is reversible, i.e. satisfies detailed balance.
Under these conditions there is a unique equilibrium measure
given by the Gibbs one
\begin{equation}
        P_{\Lambda,eq}[n]=\su{e^{-\beta
        H_\Lambda[n]}}{\sum_{\{n'\}}\,e^{-\beta H_\Lambda[n']}} ,
\end{equation} 
where $\beta \equiv 1/T$.
In the thermodynamic limit, i.e. $|\Lambda|\rightarrow \infty$, 
the lattice gas exhibits a second-order phase transition 
for $\rho_\Lambda = 1/2$ at the Onsager value
\mbox{$\beta_{c}=\su{1}{2}\ln{ (\sqrt{2}+1)}$}. 

The DLG is a generalization of the lattice gas in which one 
introduces a uniform (in space and time) force field
pointing along one of the axes of the lattice, i.e. $\ve{E}=E\ve{\hat{x}}$:
It favors (respectively suppresses) the jumps of the particles in the 
positive (resp. negative) $\ve{\hat{x}}$-direction.
If $\Lambda$\ is
bounded by {\it rigid walls}, then $\ve{E}$\ is a conservative field and 
it can be accounted for by adding a potential term
to $H_\Lambda[n]$. Therefore, the system remains in thermal equilibrium:
the net effect of $\ve{E}$ is simply to induce a concentration gradient in the
equilibrium state. 

Here, we will consider instead periodic boundary conditions \cite{foot4}.
In this case, the field $\ve{E}$ does not have a global potential and  
the system reaches a stationary state which, however, is not a state 
in thermal equilibrium.

In the \dlg\  transition
probabilities take into account the work done by the field during the
particle jump from one site to one of its nearest neighbors. In this case
one proposes a particle jump with probability 
       $w(\beta\,\Delta H+\beta E \ell )$     ,
with $\ell = (1, 0, -1)$\ for jumps (along, transverse, opposite) to 
$\ve{\hat{x}}$.

For $E \neq 0$\ and $\rho_{\Lambda}=1/2$, the system shows a continuous phase
transition at an
inverse temperature $\beta_{c}(E)$\ which saturates, for
$E\rightarrow\infty$, at $\beta_{c}(\infty) \approx 0.71
\beta_{c}(0)$. For $\beta<\beta_{c}(E)$\ particles are  homogeneously
distributed in space, while for $\beta>\beta_{c}(E)$\ the gas
separates in two regions, one almost full and the other almost empty,
with interfaces parallel to $\ve{E}$.

\subsection{Observables}
\label{sec:observables}

We consider a finite square lattice of size $\pa{L}\times
\pe{L}$ with periodic boundary conditions. We define 
a ``spin" variable $s_{\ve{j}} \equiv  2 n_{\ve{j}} - 1$ 
and its Fourier transform 
\begin{equation}
        \phi(\ve{k}) \equiv \sum_{\ve{j}\in\Lambda} 
         e^{i \ve{k}\cdot\ve{j}} s_{\ve{j}},
\end{equation}
where the allowed momenta are
\begin{equation}
        \ve{k}_{n,m} \equiv
      \left (\su{2\pi n}{\pa{L}},\su{2\pi m}{\pe{L}} \right ),
\end{equation}
with $(n,m)\in\mathbb{Z}_\protect{\pa{L}}\times\mathbb{Z}_\protect{\pe{L}} $. 

We consider the model at half filling, i.e. for $\rho_\Lambda = 1/2$. Then
\begin{equation}
        \sum_{\ve{j}\in\Lambda} s_{\ve{j}} = 0, \qquad\qquad 
        {\rm i.e.} \qquad\qquad
        \phi(\ve{k}_{0,0})=0. \label{zero}
\end{equation}
In the ordered phase $|\phi(\ve{k})|$\ takes its maximum for \mbox{$\ve{k} =
\ve{k}_{0,1}$}, and the expectation value on the steady state of its module
\begin{equation}
        m(\pa{L},\pe{L}) \equiv {1\over |\Lambda|}
            \langle|\phi(\ve{k}_{0,1})| \rangle
\label{magnetization}
\end{equation}
is a good order parameter.

In momentum space the static structure factor
\begin{equation}
        \tilde{G}(\ve{k}; \pa{L},\pe{L})\equiv 
        {1\over |\Lambda|} \langle|\phi(\ve{k})|^{2}\rangle
\label{statstrfa}
\end{equation}
vanishes at $\ve{k}_{0,0}$\ because of Eq.~(\ref{zero}) and is maximal at
$\ve{k}_{0,1}$, so that it is natural to define the susceptibility
as \cite{foot5}
\begin{equation}
        \chi (\pa{L},\pe{L}) \equiv \tilde{G}(\ve{k}_{0,1}; \pa{L},\pe{L}).
\label{defChi}
\end{equation}
Another interesting observable is the transverse Binder's 
cumulant $g(\pa{L}, \pe{L})$
defined as
\begin{equation}
\label{defbinder}
g(\pa{L}, \pe{L})  \equiv  2- 
\frac{\langle|\phi(\ve{k}_{0,1})|^{4}\rangle}
{\langle|\phi(\ve{k}_{0,1})|^{2}\rangle^2}.
\end{equation}
Next, we would like to define a correlation length. In infinite-volume 
equilibrium systems there are essentially two different ways 
of doing it. One can define the correlation length in terms of the 
large-distance behavior of the two-point function or by using the 
small-momenta behavior of the two-point function (second-moment correlation
length). In the DLG the first method does not work. 
Indeed, in the high-temperature phase the two-point function 
  $\langle s_{\ve{x}}s_{\ve{0}}\rangle$
always decays {\it algebraically} with the distance even in the
high-temperature phase. 
Moreover, it is not positive definite because of
negative correlations in the transverse directions~\cite{Schmittmann95}. 
This peculiar behavior is due to the fact
that in the infinite-volume limit (at fixed temperature)
the static structure factor 
$\tilde{G}(\ve{k};\infty,\infty)$ has a finite discontinuity at $\ve{k}=0$. 

To overcome the difficulties of the real-space strategy we will define
the correlation length
by using  the two-point function for small momenta. We follow
Ref. \cite{correlation_length}, where we discussed the possible 
definitions of correlation length in the absence of the zero mode, 
as it is the case here. 

We consider the structure factor in finite volume
at zero longitudinal momenta
\begin{equation}
        \pe{\tilde{G}}(q; \pa{L}, \pe{L}) \equiv\, 
           {\tilde{G}}((0,q); 
        \pa{L}, \pe{L}) ,
\end{equation}
(note that the conservation law 
implies $\pe{\tilde{G}}(0; \pa{L}, \pe{L})=0$)
and introduce a finite-volume (transverse) correlation length
\begin{equation}
\label{defxi}
        {\xi}_{ij}(\pa{L}, \pe{L}) \equiv
        \sqrt{\su{1}{\hat{q}_{j}^{2} - \hat{q}_{i}^{2}}
        \left( 
\su{\pe{\tilde{G}}(q_{i}; \pa{L}, \pe{L})}{
        \pe{\tilde{G}}(q_{j}; \pa{L},
        \pe{L})} -1\right) },
\end{equation}
where $\hat{q}_{n}=2\sin{(\pi n/\pe{L})}$\ is the lattice momentum.

Some comments are in order:
\begin{itemize}
\item[(i)]
If we consider an equilibrium system or a steady state in which 
correlations decay exponentially, then we have for $q\to 0$ that
\begin{equation}
        \pe{\tilde{G}}^{-1}(q; \pa{L}, \pe{L}) = 
        \chi(\pa{L}, \pe{L})^{-1}
        (  1 + {\xi}^{2}_{ij}(\pa{L}, \pe{L})\, q^{2}+ O(q^{4},L^{-2})).
\label{eq:skel}
\end{equation}
where $\chi(\pa{L}, \pe{L})$ is the susceptibility. Thus,
${\xi}^{2}_{ij}(\infty,\infty)$ is a good definition of 
correlation length which has an infinite-volume limit
independent of $i$ and $j$.

\item[(ii)] Since $\pe{\tilde{G}}(0; \pa{L}, \pe{L})=0$,
  $q_i$ and $q_j$ must not vanish. Moreover, 
  as discussed in Ref. \cite{correlation_length}, the definition
should be valid for all $T$ in finite volume. Since the system orders in an 
even number of stripes, for $i$ even $\pe{\tilde{G}}(q_i; \pa{L}, \pe{L})=0$ 
is zero as $T\to 0$. Therefore, if our definition should capture 
the nature of the phase transition, we must require $i$ and $j$ to be odd.
Although any choice of $i,j$ is conceptually good, 
finite-size corrections increase with $i,j$, a phenomenon which should 
be expected since the critical modes correspond to 
$q\to 0$. Thus, we will choose $(i,j)$ = $(1,3)$. 
\end{itemize}
For a more thorough discussion of the definition~(\ref{defxi}), which
is mainly motivated by the absence of the zero mode,  we
refer to Ref.~\cite{correlation_length} where its finite-size behavior  is
studied in the context of the large-$N$ approximation of the $N$-vector
model.

Another quantity which is considered in the analysis is the amplitude $A_{13}$
defined by
\begin{equation}
  \label{eq:amplitude}
  A_{13}(L_\|,L_\bot) \equiv  {\xi_{13}^2\over \chi}.
\end{equation}



\section{Finite-size scaling}
\label{sec:fss}

In the neighborhood of a critical point 
the behavior of long-range observables is controlled
by few quantities, corresponding in the renormalization-group language
to coordinates parametrizing the relevant
directions.
When the system is finite, its size can be effectively incorporated in
the scaling analysis as another relevant operator. 
This means that an observable
$\mathcal{O}$, which diverges in the thermodynamic limit as
\be
\mathcal{O}_\infty(\beta) \sim t^{-\gamma_\mathcal{O}} \qquad
\hbox{for } t \equiv 1-{\beta \over \beta_c} \to 0^+ ,
\end{equation}
behaves in an isotropic finite system of linear size $L$ as
\be
\mathcal{O}_L(\beta) \approx t^{-\gamma_\mathcal{O}}
F_{1,\mathcal{O}}(t^{-\nu} /L) \approx
L^{\gamma_\mathcal{O}/\nu} F_{2,\mathcal{O}}(t^{-\nu} /L) \approx 
L^{\gamma_\mathcal{O}/\nu} F_{3,\mathcal{O}}(\xi_\infty/L),
\label{FSS-basic}
\end{equation}
where $\xi_\infty(\beta)$ is the infinite-volume correlation length.
{}From Eq. \reff{FSS-basic} we can derive a general relation for
the ratio of $\mathcal{O}_L$ at two different sizes $L$ and
$\alpha L$: 
\begin{equation}
\label{eq:fssform1}
\frac{\mathcal{O}_{\alpha L}(\beta)}{\mathcal{O}_L(\beta)} = F_\mathcal{O} \left
  ( \frac{\xi_L(\beta)}{L}\right)  ,
\end{equation}
where we have traded $\xi_\infty/L$ for $\xi_L/L$ by inverting
$\xi_L \approx L F_{3,\xi}(\xi_\infty/L)$.
 
The function $F_\mathcal{O}(z)$ is universal
and is directly accessible numerically, 
e.g., by MC simulations. It depends
on $\alpha$  and, for a two-dimensional
finite lattice 
of dimension $\pe{L}\times \pa{L}$, on the ratio $\pe{L}/\pa{L}$. 
The appropriate FSS limit is indeed achieved by
taking the limit $\pe{L},\pa{L},\xi_L \to \infty$ with $\pe{L}/\pa{L}$ and 
$\xi_L/L$ fixed.    

If we define $z = \xi_L/L$, 
then $z$ varies between 0 and $z^*$, where $z^*$ is defined by 
\be
 z^* = F_{3,\xi} (\infty).
\label{def-zstar}
\end{equation}
The value $z^*$ is directly related to the behavior of the finite-size
correlation length at the critical point, since 
$\xi_L(\beta_c) \approx z^* L$. For ordinary phase transitions $z^*$ is finite.
At the critical point it also holds 
\begin{equation}
\mathcal{O}_{L}(\beta_c) \propto L^{\gamma_{\mathcal{O}}/\nu}.
\label{eq:critical_scaling}
\end{equation}
Then
\be
F_{\mathcal{O}}(z^*) =  \frac{\mathcal{O}_{\alpha
    L}(\beta_c)}{\mathcal{O}_{L}(\beta_c)} = \alpha^{\gamma_{\mathcal{O}}/\nu},
\end{equation}
and therefore
\begin{equation}
\frac{\gamma_{\mathcal{O}}}{\nu} = 
\frac{\log F_{\mathcal{O}}(z^*)}{\log \alpha}.
\label{eq:dummy_dimensions}
\end{equation}                                                                  

The above-presented results are valid for an isotropic system. 
On the other hand,
there are clear numerical evidences that the phase transition in the \dlg\ 
model is strongly anisotropic. 
The continuum field theories introduced
to describe the critical singularities predict  a nontrivial
\emph{anisotropy exponent} $\Delta$. For example, the scaling form of the
critical two-point function is predicted to be
\be
\tilde{G}(k_\bot,k_\|) \approx \mu^{-2+\eta}
\tilde{G}(\mu k_\bot,  
\mu^{1+\Delta} k_\|),
\end{equation}
where $\eta$ is the anomalous dimension of the density field. 

It is then natural to assume the existence of two correlation lengths
$\xi_\bot, \xi_\|$ which diverge with different  exponents $\nu_\bot$ and
$\nu_\|$ related by
\be
\nu_\| = (1+\Delta) \nu_\bot.
\end{equation}

These considerations call for an extension of the FSS arguments.
A phenomenological approach to \fss\ for the \dlg\ has been
developed~\cite{Valles8687}, keeping into account the strong
anisotropy observed in the transition (for $d=2$  see Refs.~\cite{Wang96,%
Leung92,Binder89}, for $d=3$ see Refs.~\cite{Leung99}).  
Following this approach, we 
assume that all observables have a finite FSS
limit obtained by taking the longitudinal size $\pa{L}$ and the
transverse size $\pe{L}$ to infinity keeping constant:
\begin{itemize}
\item the \emph{anisotropic aspect ratio} 
$
S = S_\Delta \equiv \pa{L}^{1/(1+\Delta)}/\pe{L};
$ 
\item the \emph{FSS parameter}
$\xi_{\bot,\infty}(\beta)/\pe{L}$ (or equivalently its
longitudinal counterpart).
\end{itemize}
Then, Eq. \reff{FSS-basic} still holds by using the correct FSS parameter. 
Using transverse quantities we rewrite
\be
\mathcal{O}_L(\beta) \approx t^{-\gamma_\mathcal{O}}
F_{1,\mathcal{O}}(t^{-\pe{\nu}} /\pe{L}) \approx
\pe{L}^{\gamma_\mathcal{O}/\pe{\nu}} F_{2,\mathcal{O}}(t^{-\pe{\nu}} /\pe{L}) 
    \approx
\pe{L}^{\gamma_\mathcal{O}/\pe{\nu}} 
F_{3,\mathcal{O}}(\xi_{\bot,\infty}/\pe{L}).
\label{FSS-basic-DLG}
\end{equation}
Analogously Eq.~(\ref{eq:fssform1}) is recast in the form
\begin{equation}
\label{eq:fssform2}
\frac{\mathcal{O}_{\alpha \protect{
      \pe{L}}}(\beta,S)}{\mathcal{O}_\protect{\pe{L}} (\beta,S)}
      = F_\mathcal{O} \left (
      \frac{\xi_{\bot,\pe{L}}(\beta)}{\pe{L}},\alpha, S\right).   
\end{equation}
where we have shown explicitly the dependence on $\alpha$ and on the 
aspect ratio $S$.
Equation (\ref{eq:fssform2}) will be the basis for our analysis of the phase
transition in the \dlg\ .  For the transverse finite-volume correlation 
length we will use $\xi_{13}$ as defined in Sec. \ref{sec:observables}.
Moreover, as much of our efforts are aimed to test accurately the theoretical
predictions of Refs.~\cite{Janssen86a,Leung86}, 
we \emph{assume} that $\Delta=2$ for
$d=2$ and in the sequel we always consider $S \equiv S_2$.

\section{Field-theory description of the DLG} 
\label{sec-4}

As we explained in Sec. \ref{sec-2.1}, the \dlg\ is a lattice gas
model. However, in a neighborhood of the critical
point (critical region) we can limit ourselves to consider 
slowly-varying (in space and time) observables.
At criticality the lattice spacing is
negligible compared to the length and time scales at which long-range
order is established so that it is possible to formulate a
description of the system in terms of {\it mesoscopic} variables.
In principle, the dynamics of such variables
can be obtained by coarse graining the microscopic system.
However, given the difficulty of performing a rigorous coarse-graining
procedure, one postulates a continuum field theory 
that preserves all symmetries of the microscopic lattice model. 
By universality the continuum model should have the same critical behavior 
of the microscopic one. In this 
framework a field theory has been proposed~\cite{Leung86,Janssen86a, foot2a} (see
also Ref.~\cite{Schmittmann95}) and analyzed, giving rise to exact
predictions for the critical exponents in space dimension $d$ with $2<d<5$.   

The starting point of the theory
is the dynamic functional~\cite{Janssen76,Bausch76,Janssen79} 
which reads (neglecting terms which are irrelevant by
power counting)~\cite{Janssen86a,Leung86} 
\begin{equation}
\label{thefunctional}
        J[s,\ts] = \spint{d} \
        \lambda\left\{\ts[\lambda^{-1}\partial_{t} + \nbpe(\nbpe -
        {\tau}) - \rho\nbpa]s + \su{1}{2}u_0\grpa\ts\,s^{2} +
        \ts\nbpe\ts\right\} ,
\end{equation}
where $s(\ve{x},t)$\ is the local density field (coarse-grained
version of $s_{\ve{i}}$), $\ts$\ is the response field, $\tau$\ is the
effective distance from the critical point, $\rho$\ is a parameter, and
$u_0$ is the coupling-constant of the theory which takes into account the
leading effects of the microscopic force field. 
A careful analysis shows that only the parameter $\rho$\
needs to be renormalized, allows to determine the upper critical
dimension, which turns out to be \mbox{$d_{c}=5$}, and gives exact predictions
for the critical exponents \cite{Janssen86a,Leung86}. 

For the structure factor the renormalization-group analysis gives the 
scaling form 
\begin{equation}
\tilde{G}(\pa{k},\pe{k};\tau) = \mu^{-2+\eta} 
\tilde{G}(\pa{k} \mu^{-1-\Delta},\pe{k} \mu^{-1}; \tau \mu^{-1/\nu}) ,
\label{sez4:scaling}
\end{equation}
where, in $d$ dimensions,
\begin{eqnarray}
\eta &=& 0, \\
\nu  &=& {1\over2}, \label{nu-theory} \\
\Delta &=& {1\over3} (8 - d),
\end{eqnarray}
and $\tau \propto T - T_c$. 
In two dimensions, for the transverse structure factor, this implies 
the simple scaling form 
\be
\tilde{G}_\bot(k;\tau)  = \tau^{-1} f(k^2/\tau).
\label{sec4:scaling}
\end{equation}
Thus, in infinite volume we have
\be
\xi_{ij} \sim \tau^{-\nu}, \qquad\qquad \chi \sim \tau^{-\gamma},
\label{sec4:infvol}
\end{equation}
where  $\nu$ is given in Eq. \reff{nu-theory} and 
\be
\gamma = 1.
\ee
The function $f(x)$ defined in Eq. \reff{sec4:scaling} is trivial. 
Indeed, keeping into account causality and the form of the interaction vertex
one can see that for $\pa{k} = 0$  there are no loop contributions to the 
two-point function (and also to the response function 
$\left\<\tilde{s}_0s_x\right\>$). 
Thus, for all $2\le d\le 5$, $\tilde{G}_\bot (k,\tau)$ is simply given
by the tree-level expression
\be
\tilde{G}_\bot (k;\tau) = {1\over k^2 + \tau}.
\label{sec4:tildeG}
\end{equation}
Two observation are here in order. First, $\tau$ is an analytic function 
of $t$ such that $\tau = 0$ for $t=0$; thus, $\tau = b t$ for $t\to 0$ with 
$b$ positive constant. Second, the function that appears in 
Eq. \reff{sec4:tildeG} refers to the coarse-grained fields, which, in 
the critical limit, differ by a finite renormalization from the lattice ones. 
Thus, for the lattice function we are interested in, in the scaling 
limit $t\to 0$, $k\to 0$, with $k^2/t$ fixed,
we have
\be
\tilde{G}_{\bot,\rm latt} (k;t) = {Z\over k^2 + b t},
\label{sec4:tildeG-lattice}
\ee
where $Z$ and $b$ are positive constants.

On the same footing, we can conclude that all correlation functions 
with vanishing longitudinal momenta behave as in a free
theory. In particular, the Binder cumulant defined in  Eq.~(\ref{defbinder})
vanishes. 
It is also important to notice that Eq. \reff{sec4:tildeG-lattice} implies the 
exponential decay of 
\be
    {G}_{\bot,\rm latt} (\pe{x};t) = \int d^{d-1}k \, e^{ik\pe{x}} \
      \tilde{G}_{\bot,\rm latt} (k;t), 
\end{equation}
which fully justifies our definition of transverse correlation 
length. 

In this paper, we will study the FSS behavior of the model.
Here, we want to
analyze the corresponding
continuum field theory in a finite geometry following the 
method applied in equilibrium spin systems 
(see, e.g., Ref. \cite[Chap.~36]{ZinnBook}
 and references therein). The idea  is quite simple. 
Consider the system in a finite box with periodic boundary conditions. 
The finite geometry has the only effect of quantizing the momenta. 
Thus, the perturbative finite-volume correlation functions are obtained 
by replacing momentum integrals by lattice sums. Ultraviolet divergences
are not affected by the presence of the box~\cite{Brezin82} and thus
one can use the infinite-volume renormalization constants.
Once the renormalization is carried out, one obtains the 
geometry-dependent finite-size correlation functions.

Following this idea, 
if we consider a finite box of size $\pa{L}\times \pe{L} $, 
Eq.~\reff{sez4:scaling} becomes
\be
\tilde{G}(\pa{k},\pe{k};\tau;\pa{L},\pe{L}) = \mu^{-2} 
\tilde{G}(\pa{k} \mu^{-1-\Delta},\pe{k} \mu^{-1}; \tau \mu^{-2};
   \pa{L} \mu^{-1-\Delta}, \pe{L}  \mu^{-1}) ,
\label{sec4:scaling-FSS}
\end{equation}
which shows that at $T=T_c$
\be 
\xi_{ij}(T_c) \sim \pe{L}, \qquad\qquad \chi(T_c) \sim \pe{L}^{\gamma/\nu} 
         \sim \pe{L}^2.
\label{sec4:FSS}
\end{equation}
Moreover, Eq.~\reff{sec4:tildeG} holds in finite volume. 
Keeping again into account the relation between coarse-grained 
and lattice quantities, we obtain for the lattice 
correlation function in a finite volume in the continuum limit 
(i.e. in the FSS limit with $k\to 0$ keeping $k^2/t$ fixed)
\be
    \tilde{G}_{\bot,\rm latt} (k;t;\pa{L},\pe{L}) = 
    {Z(t;\pa{L},\pe{L}) \over k^2 + \tau(t;\pa{L},\pe{L})},
\end{equation}
where $Z$ and $\tau$ are analytic functions of their arguments.
In the FSS limit, we expect
\begin{eqnarray}
     Z(t;\pa{L},\pe{L}) &=& \tilde{Z}(t \pe{L}^2, S_\Delta), \label{tildezeta}\\
     \tau(t;\pa{L},\pe{L}) &=& \pe{L}^{-2}\, \tilde{\tau}(t \pe{L}^2, S_\Delta).\label{tildetau}
\end{eqnarray}
Using these expressions, in the FSS limit we find
\begin{eqnarray}
\label{fssxi}
        \frac{{\xi}_{13}(t;\pa{L},\pe{L})}{L_\bot} &=& \left( (2\pi)^2
        + \tilde{\tau}(t \pe{L}^2, S_\Delta) \right)^{-1/2}, \\
A_{13} &\equiv&  {\xi^2_{13}(t;L_\|,L_\bot)\over \chi(t;\pa{L},\pe{L})} =  \,
           \left[\tilde{Z}(t \pe{L}^2, S_\Delta)\right]^{-1},
\end{eqnarray}
valid for $t\to 0$, $\pa{L},\pe{L}\to \infty$ with $S_2$ and 
$t \pe{L}^2$ fixed.
Therefore, from the scaling of the correlation length and of the amplitude 
$A_{13}$ we can derive the scaling functions $\tilde{Z}$ and $\tilde{\tau}$.

If we make the simplest approximations 
$\tilde{Z}(t \pe{L}^2, S_\Delta) = {\rm const}$ and 
$\tilde{\tau}(t \pe{L}^2, S_\Delta) = {\rm const}\  \times t \pe{L}^2$, 
we obtain 
for the scaling functions defined in
Eq. \reff{eq:fssform2} the approximate forms
\begin{eqnarray}
        F_{\xi}(z) &=& \left[ 1- \left(1 -\alpha^{-2}\right)(2\pi)^2 z^2
         \right]^{-1/2},
\label{eq:theoretical_pred} \\
F_{\chi} (z) &=& F_\xi^2(z) = 
   \left[1- \left(1 -\alpha^{-2}\right)(2\pi)^2 z^2\right]^{-1}.
\label{eq:theoretical_pred_chi} 
\end{eqnarray}
As we shall see, these expressions provide reasonably good 
approximations to our data, but do not describe the data exactly: The 
functions $\tilde{\tau}$ and $\tilde{Z}$ that are determined from the data 
are nontrivial.

In the previous discussion
we have neglected the possible presence of a dangerously irrelevant 
operator that becomes marginal at $d=2$. Its presence may modify
the scaling relations \reff{sez4:scaling}, \reff{sec4:scaling},
and \reff{sec4:scaling-FSS}. Considering $\pe{\tilde{G}} (k)$ we 
have
\be
\pe{\tilde{G}} (k; \tau; \pa{L}, \pe{L}; u) = 
   \mu^{-2} \pe{\tilde{G}} \left(k \mu^{-1}; \tau \mu^{-2}; 
     \pa{L} \mu^{-1-\Delta}, \pe{L}\mu^{-1}; u \mu^{2\sigma} \right),
\ee
where, $\sigma = (d-2)/3$ in $d$ dimensions, and $u$ is the irrelevant 
coupling. If the scaling function vanishes for $u\to 0$, one obtains
an anomalous scaling behavior. In two dimensions, since the operator is 
marginal ($\sigma=0$), we expect logarithmic corrections to the 
formulae previously computed.
It is also possible that logarithmic terms modify 
Eq. \reff{eq:fssform2}. In the absence of any prediction, we will neglect
these logarithmic violations. As it has been observed in previous 
numerical studies, if present, they are small \cite{Leung92}. 
As we will discuss, this is confirmed by our numerical results.

Finally, we want to notice that the Gaussian nature of the fields 
at zero longitudinal momenta allow us to write down the 
probability distribution of $\phi({\bf k}_{0,n})$. 
Indeed, the previous results give 
\be
P(\phi({\bf k}_{0,n})) = {1\over N} \exp\left[- {1\over \pa{L} \pe{L}} 
   \phi(-{\bf k}_{0,n})  \tilde{G}^{-1}_{\bot,\rm latt}({\bf k}_{0,n}) 
   \phi({\bf k}_{0,n} )\right],
\ee
where $N$ is a normalization factor. From this expression we can 
derive the probability distribution of 
$M^2 = |\phi({\bf k}_{0,1})|^2/\pa{L}^2 \pe{L}^2$. 
Of course, $\<M\> = m$, $\<M^2\> = \chi/(\pa{L} \pe{L})$\ 
(see Eqs.~\reff{magnetization}, \reff{statstrfa}, and \reff{defChi}). We obtain
\be
P(M^2) dM^2 = {1\over \sigma^2} e^{-M^2/\sigma^2} dM^2,
\label{prediction-M2-distribution}
\ee
where 
\be 
\sigma^2 = {1\over \pa{L} \pe{L}} \tilde{G}_{\bot,\rm latt}(2\pi/L_\bot) = 
    {1\over \pa{L} \pe{L}} \chi(\pa{L}, \pe{L}).
\ee

\section{Numerical Simulation}
\label{sec:simulations}

\subsection{Setup}

We studied the phase transition of the \dlg\ in two dimensions by 
MC simulations. Our aim was to check 
the validity of the FSS assumptions and eventually to use FSS to
compute the critical exponents. The observables we took into account
are those described in Sec.~\ref{sec:observables}.

We used the dynamics described in Sec.~\ref{sec-2.1} with Metropolis rates,
i.e. we set
\be
w(x) =\, {\rm min}\ (1,e^{-x}).
\end{equation}
Simulations are performed at infinite driving field: therefore,
forward (backward) jumps in the direction of the field are always
accepted (rejected).

The dynamics of the \dlg\ is diffusive with dynamic critical 
exponent \cite{Janssen86a} $z=4$. Thus, it is important to have
an efficient implementation of the dynamics in order to cope with 
the severe critical slowing down.     

We use the random-number generator {\tt  ran\_array} 
described in Ref.~\cite{knuth} in order to choose the
links and  to perform the Metropolis step
and multi-spin coding techniques to evolve simultaneously many
independent configurations. On 32-bit machines the state of a single
site for 10 configurations is coded in a 32-bit word, using 3 bits per
configuration. This peculiar allocation is useful when
computing the number of occupied neighbors of a given site: Since this
number is at most 4 in 2 dimensions we can use 32-bit arithmetics
without overflow.
The Metropolis step is coded in an entirely
algebraic way avoiding branching instructions and taking full
advantage of the pipelining capabilities of most microprocessors. 
On Pentium machines we achieved a speed of 0.01 $\mu$sec for
single spin flip,  while our code performs a bit worse on Motorola
$R10000$ or Alpha processors but on these machines it is possible to
simulate $21$ systems using $64$-bit integers.

We considered lattices with an approximately constant aspect ratio 
$S_2\approx 0.200$.
We sampled very accurately those with sizes $(L_\|,L_\bot)$ corresponding to
$(21,14)$, $(32,16)$, $(46,18)$, $(64,20)$, $(84,22)$, $(110,24)$, $(168,28)$
for which $S_2$ lies between $0.197$ and $0.202$. We checked that
a change of $\pm1$ on $L_\|$ has an effect on the observables that is 
negligible
with respect to the statistical error.  
In the following we write $L$ for $L_\bot$.      

It is very important to be sure that the system has reached the steady-state
distribution before sampling. 
Metastable configurations in which the chain could be trapped for times
much longer than typical relaxation times in the steady state are
a dangerous source of bias. In the \dlg ,
configurations in which multiple stripes aligned with the external
field are present are very long-lived and it is possible that
they persist for times of the order of those of typical simulation runs, thus
effectively inducing a spurious geometry on the system. To avoid the
formation of stripes, we took care to
initialize the larger systems by suitably rescaled thermalized configurations
of smaller systems (where the stripes decay faster) at the same temperature
and value of $S_2$.  

We computed the autocorrelation time $\tau_\chi$ for the susceptibility $\chi$. 
Such an observable is expected to have a significant overlap with the slowest 
modes of the system, so that $\tau_\chi$ should give a good indication 
of the number of sweeps necessary to generate independent configurations.
We found that, for the lowest temperature we considered ($\beta=0.311$), 
$\tau_\chi \approx 900$, 1400, $2700$  sweeps, for $L=20,24,28$
respectively,  where a MC sweep is 
conventionally defined as the number of moves
equal to the volume of the lattice. For this reason, in order to have 
approximately independent configurations, we measured once every 
$1500$ sweeps. 

For each geometry, we performed long simulations 
at $15$ different values of $\beta$ in the disordered phase
between $0.27$ and $0.311$, which correspond to temperatures ranging 
from $1.41654$ to
$1.63217$ in units of the critical temperature for the 
two-dimensional Ising model $\beta_c
= \log(1+\sqrt{2})/2 \approx 0.440687$.  
For each geometry and $\beta$ we collected approximately $10^5$ measures.
The raw data are reported in Tables \ref{table_tot_1} and \ref{table_tot_2}.

\begin{table}[t]
\begin{center}
{
\tiny
\begin{tabular}{|l|l|l|l|l|l|l|}
 \hline 
 \hfil $L$ \hfil &  \hfil $\xi_{12}$ \hfil & \hfil $\xi_{13}$ \hfil & \hfil 
  $\chi$ \hfil & \hfil $A_{13}$ \hfil & \hfil $m$ \hfil & \hfil $g$ \hfil  \\ 
 \hline 
 & \multicolumn{6}{|c|}{$ \beta = 0.2700 $} \\ 
 \hline 
14 & 1.3824(16) & 1.3769(11) & 8.0489(50) & 0.23555(28) & 0.148869(50) & 0.20672(73)\\ 
16 & 1.4455(21) & 1.4448(15) & 8.8606(69) & 0.23558(36) & 0.117807(49) & 0.1495(10)\\ 
18 & 1.4942(27) & 1.4905(17) & 9.4795(83) & 0.23436(41) & 0.095533(44) & 0.1105(10)\\ 
20 & 1.5294(35) & 1.5283(22) & 9.960(10) & 0.23450(51) & 0.078584(43) &
0.0786(13)\\ 
\hline 
  & \multicolumn{6}{|c|}{$ \beta = 0.2750 $} \\ 
 \hline 
14 & 1.4532(16) & 1.4455(11) & 8.7094(53) & 0.23992(27) & 0.155093(51) & 0.22560(72)\\ 
16 & 1.5252(22) & 1.5206(15) & 9.6826(74) & 0.23881(35) & 0.123347(51) & 0.16947(98)\\ 
18 & 1.5793(28) & 1.5796(18) & 10.4511(93) & 0.23873(41) & 0.100445(47) & 0.1273(11)\\ 
20 & 1.6280(35) & 1.6255(22) & 11.073(12) & 0.23862(49) & 0.082940(46) & 0.0923(13)\\ 
\hline 
  & \multicolumn{6}{|c|}{$ \beta = 0.2800 $} \\ 
 \hline 
14 & 1.5286(16) & 1.5172(11) & 9.4382(56) & 0.24389(27) & 0.161823(52) & 0.25090(68)\\ 
16 & 1.6159(21) & 1.6118(15) & 10.6473(81) & 0.24399(34) & 0.129594(53) & 0.19095(95)\\ 
18 & 1.6858(26) & 1.6821(19) & 11.641(10) & 0.24304(40) & 0.106178(49) & 0.1473(10)\\ 
20 & 1.7428(34) & 1.7408(22) & 12.421(13) & 0.24396(46) & 0.087965(48) & 0.1101(12)\\ 
\hline 
  & \multicolumn{6}{|c|}{$ \beta = 0.2850 $} \\ 
 \hline 
14 & 1.6086(16) & 1.5940(11) & 10.2434(60) & 0.24803(26) & 0.168977(54) & 0.27700(65)\\ 
16 & 1.7152(21) & 1.7048(15) & 11.7579(87) & 0.24719(33) & 0.136493(54) & 0.21965(88)\\ 
18 & 1.7998(27) & 1.7963(19) & 13.029(11) & 0.24765(38) & 0.112516(51) & 0.16941(98)\\ 
20 & 1.8654(34) & 1.8657(23) & 14.067(15) & 0.24745(45) & 0.093766(51) & 0.1334(12)\\ 
\hline 
  & \multicolumn{6}{|c|}{$ \beta = 0.2900 $} \\ 
 \hline 
14 & 1.6976(16) & 1.6769(12) & 11.1664(64) & 0.25183(26) & 0.176906(55) & 0.30542(61)\\ 
16 & 1.8243(22) & 1.8130(16) & 13.0604(95) & 0.25168(33) & 0.144208(57) & 0.24750(84)\\ 
18 & 1.9347(27) & 1.9256(19) & 14.729(12) & 0.25174(37) & 0.119924(54) & 0.20021(91)\\ 
20 & 2.0225(34) & 2.0137(24) & 16.134(17) & 0.25132(43) & 0.100638(55) & 0.1608(11)\\ 
22 & 2.0904(31) & 2.0862(21) & 17.310(15) & 0.25142(38) & 0.086542(40) & 0.1292(11)\\ 
24 & 2.1457(35) & 2.1447(23) & 18.304(17) & 0.25129(39) & 0.074305(35) & 0.1026(12)\\ 
28 & 2.2444(45) & 2.2389(29) & 19.900(22) & 0.25190(46) & 0.057904(33) & 0.0685(14)\\ 
\hline 
  & \multicolumn{6}{|c|}{$ \beta = 0.2950 $} \\ 
 \hline 
14 & 1.7941(16) & 1.7685(12) & 12.2087(68) & 0.25617(26) & 0.185533(57) & 0.33682(58)\\ 
16 & 1.9527(22) & 1.9338(16) & 14.602(10) & 0.25609(32) & 0.152967(58) & 0.28268(78)\\ 
18 & 2.0893(27) & 2.0740(20) & 16.797(14) & 0.25609(37) & 0.128498(56) & 0.23771(85)\\ 
20 & 2.2041(34) & 2.1941(25) & 18.754(19) & 0.25670(41) & 0.108808(60) & 0.1956(10)\\ 
22 & 2.2962(32) & 2.2878(22) & 20.434(18) & 0.25614(36) & 0.094271(43) & 0.1623(10)\\ 
24 & 2.3770(33) & 2.3731(23) & 21.937(19) & 0.25671(35) & 0.081524(36) & 0.1315(10)\\ 
28 & 2.5045(50) & 2.5028(33) & 24.333(29) & 0.25743(46) & 0.064138(40) & 0.0922(14)\\ 
\hline 
  & \multicolumn{6}{|c|}{$ \beta = 0.2970 $} \\ 
 \hline 
14 & 1.8368(16) & 1.8072(12) & 12.6676(70) & 0.25783(26) & 0.189260(58) & 0.35048(57)\\ 
16 & 2.0076(22) & 1.9852(16) & 15.298(11) & 0.25762(32) & 0.156795(60) & 0.29750(76)\\ 
18 & 2.1538(27) & 2.1369(20) & 17.709(15) & 0.25786(35) & 0.132092(58) & 0.25094(84)\\ 
20 & 2.2786(36) & 2.2698(26) & 19.949(21) & 0.25826(41) & 0.112359(62) & 0.2100(10)\\ 
22 & 2.3867(32) & 2.3778(23) & 21.940(19) & 0.25771(35) & 0.097791(45) & 0.1773(10)\\ 
24 & 2.4838(33) & 2.4787(24) & 23.735(20) & 0.25885(34) & 0.084924(39) & 0.1488(10)\\ 
28 & 2.6308(53) & 2.6194(36) & 26.590(34) & 0.25804(46) & 0.067096(45) & 0.1030(14)\\ 
\hline 
\end{tabular}
}
\end{center}
\caption{Monte Carlo results.}
\label{table_tot_1}
\end{table}

\begin{table}
\begin{center}
{
\tiny
\begin{tabular}{|l|l|l|l|l|l|l|}
 \hline 
 \hfil $L$ \hfil &  \hfil $\xi_{12}$ \hfil & \hfil $\xi_{13}$ \hfil & \hfil 
  $\chi$ \hfil & \hfil $A_{13}$ \hfil & \hfil $m$ \hfil & \hfil $g$ \hfil  \\ 
 \hline 
  & \multicolumn{6}{|c|}{$ \beta = 0.3075 $} \\ 
 \hline 
14 & 2.0812(17) & 2.0241(12) & 15.4022(79) & 0.26599(25) & 0.210342(60) & 0.42478(49)\\ 
16 & 2.3396(23) & 2.2947(17) & 19.719(13) & 0.26703(30) & 0.179677(65) & 0.38902(64)\\ 
18 & 2.5907(28) & 2.5477(22) & 24.273(19) & 0.26740(34) & 0.156237(65) & 0.35715(68)\\ 
20 & 2.8299(40) & 2.7938(30) & 29.116(31) & 0.26807(38) & 0.137182(79) & 0.32443(87)\\ 
22 & 3.0510(33) & 3.0174(25) & 33.872(28) & 0.26880(31) & 0.122863(55) & 0.30119(79)\\ 
24 & 3.2719(39) & 3.2349(29) & 38.911(36) & 0.26894(31) & 0.109908(55) & 0.27470(86)\\ 
28 & 3.6789(65) & 3.6355(49) & 48.776(74) & 0.27097(40) & 0.091858(75) & 0.2360(13)\\ 
\hline 
  & \multicolumn{6}{|c|}{$ \beta = 0.3085 $} \\ 
 \hline 
14 & 2.1082(14) & 2.0488(11) & 15.7157(69) & 0.26710(22) & 0.212671(52) & 0.43320(41)\\ 
16 & 2.3781(16) & 2.3270(12) & 20.2393(93) & 0.26754(22) & 0.182213(46) & 0.39818(45)\\ 
18 & 2.6418(18) & 2.5947(14) & 25.103(12) & 0.26819(22) & 0.159092(42) & 0.36907(47)\\ 
20 & 2.8958(21) & 2.8542(16) & 30.281(16) & 0.26903(22) & 0.140120(40) & 0.34012(51)\\ 
22 & 3.1328(36) & 3.0961(27) & 35.514(31) & 0.26992(33) & 0.125993(60) & 0.31663(82)\\ 
24 & 3.3744(39) & 3.3379(30) & 41.131(38) & 0.27089(31) & 0.113183(58) & 0.29252(82)\\ 
28 & 3.8316(69) & 3.7822(51) & 52.382(80) & 0.27308(40) & 0.095379(79) & 0.2569(13)\\ 
\hline 
  & \multicolumn{6}{|c|}{$ \beta = 0.3000 $} \\ 
 \hline 
14 & 1.8998(16) & 1.8659(12) & 13.3718(72) & 0.26038(25) & 0.194851(58) & 0.37052(54)\\ 
16 & 2.0932(22) & 2.0671(17) & 16.418(11) & 0.26026(31) & 0.162810(61) & 0.32164(73)\\ 
18 & 2.2586(27) & 2.2420(20) & 19.288(15) & 0.26061(35) & 0.138209(59) & 0.27788(78)\\ 
20 & 2.4132(36) & 2.3997(27) & 22.069(23) & 0.26093(40) & 0.118453(66) & 0.23766(98)\\ 
22 & 2.5425(33) & 2.5327(24) & 24.588(21) & 0.26088(34) & 0.103782(48) & 0.20613(97)\\ 
24 & 2.6602(34) & 2.6487(24) & 26.882(23) & 0.26098(32) & 0.090555(42) & 0.17492(97)\\ 
28 & 2.8636(55) & 2.8495(39) & 30.910(41) & 0.26269(44) & 0.072488(51) & 0.1300(14)\\ 
\hline 
  & \multicolumn{6}{|c|}{$ \beta = 0.3025 $} \\ 
 \hline 
14 & 1.9568(16) & 1.9155(12) & 14.0015(73) & 0.26205(25) & 0.199740(58) & 0.38791(52)\\ 
16 & 2.1682(22) & 2.1408(17) & 17.417(12) & 0.26314(31) & 0.168038(62) & 0.34245(70)\\ 
18 & 2.3616(27) & 2.3359(21) & 20.789(16) & 0.26247(34) & 0.143824(61) & 0.30275(75)\\ 
20 & 2.5389(37) & 2.5164(28) & 24.106(25) & 0.26269(40) & 0.124114(69) & 0.26489(95)\\ 
22 & 2.6910(33) & 2.6756(25) & 27.182(24) & 0.26336(34) & 0.109385(51) & 0.23325(95)\\ 
24 & 2.8355(35) & 2.8220(26) & 30.160(27) & 0.26404(32) & 0.096155(46) & 0.20347(94)\\ 
28 & 3.0806(57) & 3.0613(41) & 35.373(49) & 0.26494(42) & 0.077706(57) & 0.1565(14)\\ 
\hline 
  & \multicolumn{6}{|c|}{$ \beta = 0.3050 $} \\ 
 \hline 
14 & 2.0171(16) & 1.9683(12) & 14.6780(77) & 0.26395(25) & 0.204930(60) & 0.40679(51)\\ 
16 & 2.2498(23) & 2.2116(17) & 18.515(12) & 0.26418(31) & 0.173660(64) & 0.36476(68)\\ 
18 & 2.4701(28) & 2.4387(21) & 22.436(17) & 0.26508(34) & 0.149777(63) & 0.32770(72)\\ 
20 & 2.6757(37) & 2.6500(28) & 26.388(27) & 0.26613(38) & 0.130181(72) & 0.29203(89)\\ 
22 & 2.8641(34) & 2.8364(26) & 30.272(26) & 0.26577(33) & 0.115770(55) & 0.26642(88)\\ 
24 & 3.0339(36) & 3.0105(27) & 34.042(30) & 0.26624(32) & 0.102440(49) & 0.23553(89)\\ 
28 & 3.3461(65) & 3.3163(48) & 41.058(64) & 0.26786(44) & 0.083931(70) & 0.1886(14)\\ 
\hline 
  & \multicolumn{6}{|c|}{$ \beta = 0.3100 $} \\ 
 \hline 
14 & 2.1460(12) & 2.08097(94) & 16.1664(61) & 0.26787(19) & 0.215981(45) & 0.44418(35)\\ 
16 & 2.4343(12) & 2.37889(90) & 21.0314(69) & 0.26908(16) & 0.186047(34) & 0.41292(31)\\ 
18 & 2.7193(15) & 2.6677(11) & 26.3595(100) & 0.26998(17) & 0.163321(34) & 0.38646(37)\\ 
20 & 3.0030(17) & 2.9520(13) & 32.194(14) & 0.27068(18) & 0.144787(34) & 0.36129(40)\\ 
22 & 3.2738(37) & 3.2257(28) & 38.277(33) & 0.27184(33) & 0.131133(62) & 0.34229(79)\\ 
24 & 3.5552(41) & 3.5032(31) & 44.923(42) & 0.27319(31) & 0.118619(61) & 0.32141(81)\\ 
28 & 4.0953(72) & 4.0294(55) & 58.956(93) & 0.27539(40) & 0.101535(88) & 0.2944(13)\\ 
\hline 
  & \multicolumn{6}{|c|}{$ \beta = 0.3105 $} \\ 
 \hline 
14 & 2.1577(15) & 2.0938(11) & 16.3170(75) & 0.26867(23) & 0.217091(56) & 0.44857(42)\\ 
16 & 2.4548(17) & 2.3945(13) & 21.297(10) & 0.26921(22) & 0.187326(49) & 0.41820(45)\\ 
18 & 2.7459(20) & 2.6913(15) & 26.792(13) & 0.27035(23) & 0.164752(45) & 0.39204(48)\\ 
20 & 3.0396(23) & 2.9862(17) & 32.881(18) & 0.27121(23) & 0.146452(44) & 0.36941(52)\\ 
22 & 3.3226(37) & 3.2667(28) & 39.201(34) & 0.27222(33) & 0.132833(63) & 0.35071(79)\\ 
24 & 3.6092(41) & 3.5544(31) & 46.245(43) & 0.27320(31) & 0.120487(61) & 0.33219(80)\\ 
28 & 4.1856(73) & 4.1161(55) & 61.352(96) & 0.27615(39) & 0.103692(89) & 0.3059(12)\\ 
\hline 
  & \multicolumn{6}{|c|}{$ \beta = 0.3110 $} \\ 
 \hline 
14 & 2.1746(12) & 2.10673(94) & 16.4952(62) & 0.26906(18) & 0.218364(46) & 0.45238(35)\\ 
16 & 2.4750(17) & 2.4135(13) & 21.591(10) & 0.26980(23) & 0.188723(49) & 0.42338(45)\\ 
18 & 2.7757(20) & 2.7185(15) & 27.273(13) & 0.27097(23) & 0.166366(45) & 0.39995(47)\\ 
20 & 3.0736(23) & 3.0208(18) & 33.572(18) & 0.27182(23) & 0.148113(44) & 0.37754(51)\\ 
22 & 3.3632(38) & 3.3148(29) & 40.241(35) & 0.27305(32) & 0.134690(64) & 0.35948(77)\\ 
24 & 3.6772(41) & 3.6188(32) & 47.716(44) & 0.27446(30) & 0.122489(63) & 0.34152(78)\\ 
28 & 4.3024(76) & 4.2147(58) & 64.14(10) & 0.27697(39) & 0.106167(93) & 0.3206(12)\\ 
\hline 
 \end{tabular}
}
\end{center}
\caption{Monte Carlo results.}
\label{table_tot_2}
\end{table}

The statistical variance of the observables is estimated by using the jackknife
method~\cite{Efron82}.   
To take into account the  possible residual correlations of the samples,
we used a blocking technique in the jackknife analysis. In the standard
jackknife method the estimator for the variance is obtained discarding
single data points. In the blocking technique, several variance estimators
are considered, discarding blocks of data of increasing length and monitoring 
the estimated variance until it reaches a maximum.

\subsection{Finite-size scaling and extrapolation to the infinite-volume limit}
\label{sec5:FSS}

In this Section we want to test the validity of Eq.~(\ref{eq:fssform2}),
by studying the FSS of the observables defined in Sec. \ref{sec-2}. 
In order to compute $F_{\cal O}(z)$, we need to keep the ratio
between the transverse sizes of two different lattices equal to
$\alpha$. The simplest thing to do would be to take $\alpha=2$ but
this choice has the drawback of requiring large systems: indeed, 
if $\pe{L}$ increases by a factor of two, $\pa{L}$ increases by a 
factor of $2^{1+\Delta} = 8$. A more
convenient choice consists in taking a noninteger 
$\alpha$ almost equal to $1$, as far
as it is allowed by the magnitude of the relative statistical errors. 
This strategy requires the knowledge of the values of the observables for
noninteger values of $L_\bot$. They are obtained by using an
interpolation.  In the analysis we used  $\alpha=1.25$.

At each $\beta=\beta_a$, $a=1,\dots,n$, our data consist of $n^a$ triples 
$(L^a_{i},\xi^a_{i},\obs^a_{i})$, $i=1,\dots,n^a$, of lattices and observed
values. 
The data points are interpolated with smooth cubic splines 
$\widehat{\xi}^a(L)$, $\widehat{\obs}^a(L)$. This interpolation 
provides estimates of $\xi^a(L)$ and ${\obs}^a(L)$ for any
(not necessarily integer) value of $L$ between 
$L^a_{\rm min} = {\rm min}_i\ L_i^a$ and 
$L^a_{\rm max} = {\rm max}_i\ L_i^a$. Then, we consider a set of 
$m^a \equiv \lfloor (n^a+1)/2 \rfloor$ 
values of $L$,    
\be
\tilde{L}^a_j = L^a_{\rm min} + {(j-1)\over (m^a - 1)} 
     \left(\alpha^{-1} L^a_{\rm max} - L^a_{\rm min}\right),
\end{equation}
where $j=1,\ldots,m^a$, and compute the corresponding triples
\be
A^a_j \equiv (z^a_j,R^a_{\xi,j},R^a_{\obs,j}) =
\left(
  \frac{\widehat{\xi}^a(\tilde L^a_j)}{\tilde L^a_j},
  \frac{\widehat{\xi}^a(\alpha \tilde L^a_j)}{\widehat{\xi}^a(\tilde L^a_j)},
  \frac{\widehat{\obs}^a(\alpha\tilde L^a_j)}{\widehat{\obs}^a(\tilde L^a_j)}
\right)
\label{triples-A}
\end{equation}
for $j=1,\ldots,m^a$. The triples \reff{triples-A} are the data we use in 
the computation of the FSS functions. Note
that, since we need two independent data at different values of $L$ to
compute $R$, we generate a number of interpolated points
that is one half of the total amount of data. 
This minimizes the correlations among the $A^a_j$. 

Errors are computed by using an auxiliary MC procedure that also estimates 
the covariance matrices $(C^a_{\xi})_{i j}=\cov(R^a_{\xi,i},R^a_{\xi,j})$ and 
$(C^a_\obs)_{i j} = \cov(R^a_{\obs,i},R^a_{\obs,j})$.    

The points $(z^a_j,R^a_{\obs,j})$ are fitted
to $(z,\hat F_\obs(z))$ where $\hat F_\obs(z)$ is a 
polynomial in $z$ of the form
\be
\hat F_\obs (z) = 1 + a_{\obs,1}\, z^2 + 
a_{\obs,2}\, z^3 + a_{\obs,3}\, z^4 + \cdots 
\label{fitting-function}
\end{equation}
Such a parametrization, different from that used in Ref. \cite{Caracciolo95},
is partially motivated by the theoretical results of Sec.~\ref{sec-4}, 
indicating 
that, for small $z$, the FSS functions behave as $1 + a z^2$.
If we neglect the correlations between $z_j^a$ and
$R^a_{\obs,j}$, then the $\chi^2$ minimization is a linear
problem which is easily solvable \cite{foot6}.

In order to detect corrections to scaling, we perform the fit several times,
including in each case only the results that satisfy 
$\tilde{L}^a_j \ge L_{\text{min},\obs}$, increasing step-by-step the 
parameter $L_{\text{min},\obs}$. 

\begin{table}
\begin{center}
{
\scriptsize
\begin{tabular}{|c||c|c||c|c||c|c||c|c||}
\hline
 & 
\multicolumn{2}{|c|}{$L_{\text{min}} = 15.4$} &
\multicolumn{2}{|c|}{$L_{\text{min}} = 16.8$} &
\multicolumn{2}{|c|}{$L_{\text{min}} = 18.2$} &
\multicolumn{2}{|c|}{$L_{\text{min}} = 21$} \\
\hline
$\beta$ 
& $\xi_{\infty}$ & $R^2(\text{DOF})$ 
& $\xi_{\infty}$ & $R^2(\text{DOF})$ 
& $\xi_{\infty}$ & $R^2(\text{DOF})$ 
& $\xi_{\infty}$ & $R^2(\text{DOF})$ \\
\hline

0.27     & 1.6703(69) &   1.2(2)         & 1.6671(82) &   0.4(1)         & 1.680(12) &   0.0(0)  &  &   \\
0.275    & 1.8085(84) &   0.5(2)         & 1.8042(95) &   0.3(1)         & 1.820(13) &   0.0(0)  &  &   \\
0.28     & 1.9827(98) &   0.9(2)         & 1.979(11) &   1.0(1)  & 1.996(17) &   0.0(0)  &  &   \\
0.285    & 2.194(11) &   2.1(2)  & 2.189(13) &   0.1(1)  & 2.207(20) & 0.0(0)  &  &   \\
0.29     & 2.477(11) &   3.1(5)  & 2.470(13) &   3.2(4)  & 2.495(17) &   1.3(3)  & 2.502(25) &   0.7(2)\\
0.295    & 2.884(14) &   3.8(5)  & 2.876(17) &   2.8(4)  & 2.901(24) &   2.2(3)  & 2.914(32) &   0.3(2)\\
0.297    & 3.090(15) &   6.1(5)  & 3.082(18) &   5.8(4)  & 3.101(26) &   6.2(3)  & 3.108(37) &   6.4(2)\\
0.3      & 3.521(18) &   3.9(5)  & 3.513(21) &   4.0(4)  & 3.545(33) &   3.6(3)  & 3.570(44) &   2.4(2)\\
0.3025   & 4.026(21) &   1.5(5)  & 4.012(25) &   0.9(4)  & 4.042(38) &   0.9(3)  & 4.069(48) &   0.6(2)\\
0.305    & 4.793(25) &   10(5)         & 4.790(30) &   4.6(4)  & 4.816(41) &   1.9(3)  & 4.854(58) &   0.4(2)\\
0.3075   & 6.269(33) &   3.4(5)  & 6.257(38) &   3.4(4)  & 6.256(52) &   2.4(3)  & 6.332(72) &   0.05(2)\\
0.3085   & 7.391(38) &   7.7(5)  & 7.390(42) &   4.7(4)  & 7.357(58) &   2.9(3)  & 7.464(85) &   0.7(2)\\
0.31     & 11.506(94) &   7.7(5)         & 11.49(10) &   4.5(4)  & 11.22(13) &   2.0(3)  & 11.60(20) &   1.3(2)\\
0.3105   & 15.14(23) &   12.5(5)         & 15.21(24) &   9.2(4)  & 14.55(28) &   5.1(3)  & 15.55(48) &   0.1(2)\\
\hline

\end{tabular}
}
\end{center}
\caption{Infinite-volume results for $\xi_{13}$, obtained from the 
extrapolation of finite-size data, for different values of 
  $L>L_{\text{min}}$. $R^2$ is the $\chi^2$ defined in Eq.
  \reff{R2}, and DOF is the number of degrees of freedom.}
\label{table:extr_xi}
\end{table}

The number $N_{\text{par},\obs}$ of parameters in the fit as well as 
the value of $L_{\text{min},\obs}$ are chosen in such a way to optimize the 
values of the corresponding $\chi^2$. 
In particular, at fixed $L_{\text{min},\obs}$, we take the value of 
$N_{\text{par},\obs}$ which minimizes 
$\chi^2/N_{\text{dof},\obs}$, where $N_{\text{dof},\obs}$
is the number of degrees of freedom of the fit (the number of data
points minus $N_{\text{par},\obs}$); then, we increase $L_{\text{min},\obs}$
until we attain $\chi^2/N_{\text{dof},\obs} \approx 1$ and 
the curve is compatible with the points with the largest value of $L$.
   
\begin{figure}
\vspace*{-2.0truecm}
\begin{center}
\epsfig{file=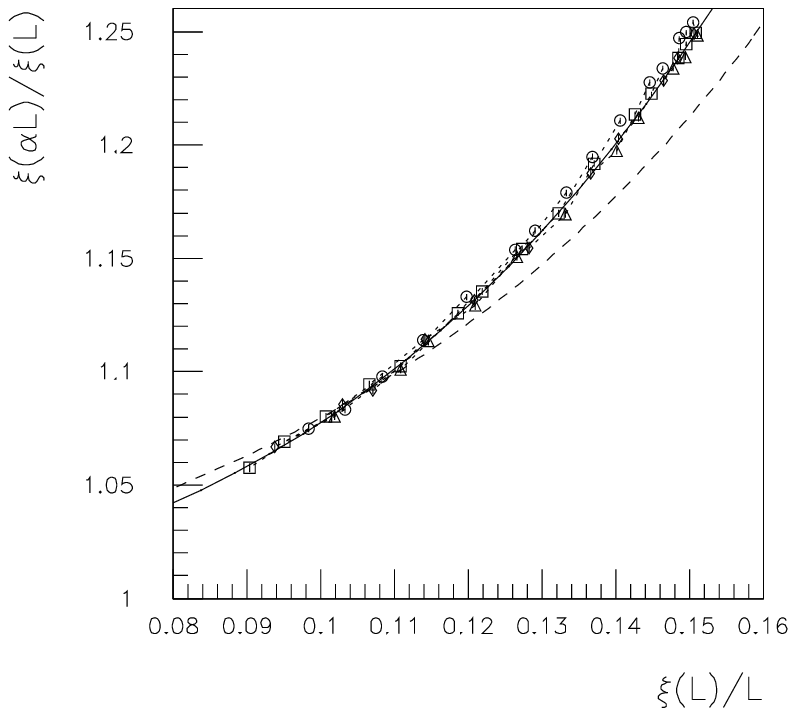,width=0.7\textwidth}
\end{center}
\vspace*{-1.5truecm}
\caption{FSS plot for $\xi_{13}$ versus $\xi_{13}$. 
The symbols correspond to the 
following lattice sizes: $(\circ)$ $ L\le 15$, $( \square )$ $ 15 < L \le 17$,
$(\triangle)$ $ 17 < L \le 22$, $(\lozenge)$ $ L > 22$. 
Dotted lines connect data at the same temperatures.
The dashed line is the 
approximation \reff{eq:theoretical_pred}, the continuous line is the 
result of the fit with $L_{\rm min} = 15.4$.}
\label{fig:fssplot_xi13}
\end{figure}

\begin{figure}
\begin{center}
\epsfig{file=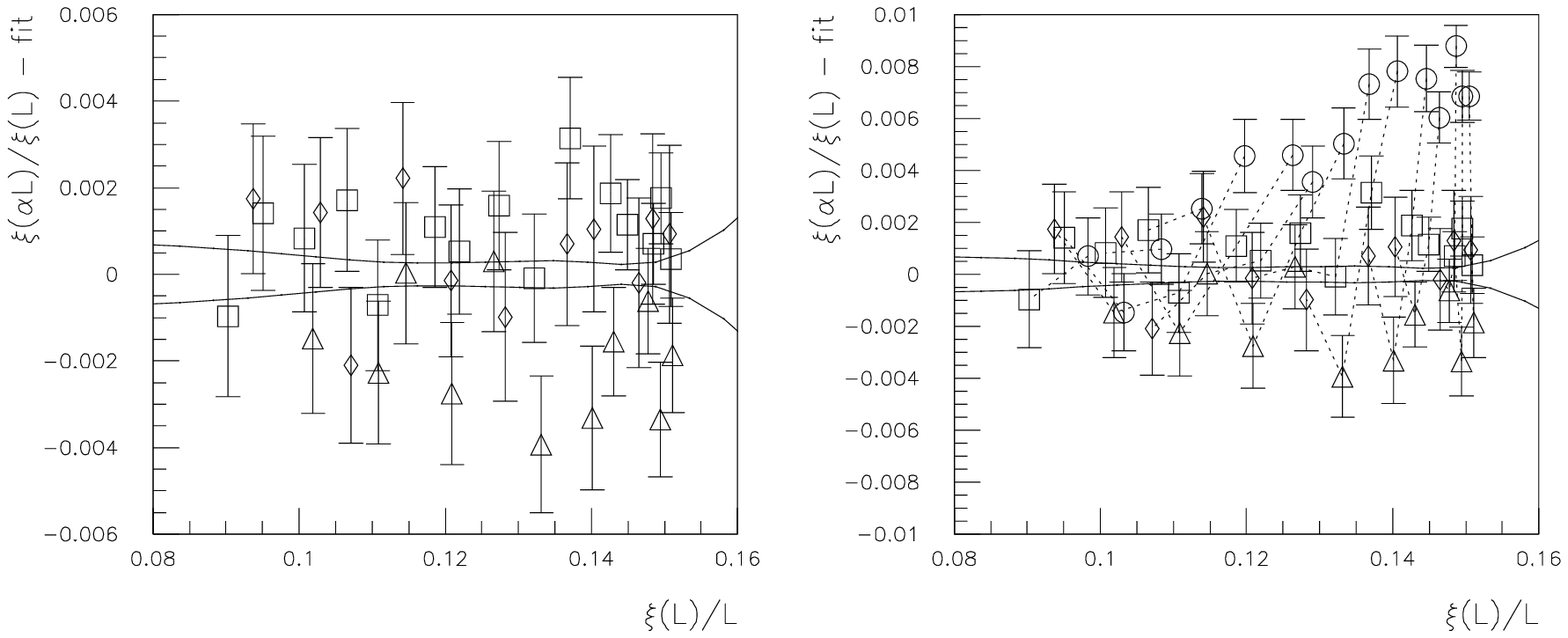,width=0.8\textwidth}
{\scriptsize
\begin{tabular}{|c|c|c|c|c|}
\hline
$L_{\text{min}}$ & $N_{\text{par}}$ & $\chi^2$ & $N_{\text{dof}}$ &
$\chi^2/N_{\text{dof}}$ \\
\hline
 14    &     3  &      286 &  49 &     5.8\\
 15.4  &     3  &      39 &  34 &     1.1\\
 16.8  &     3  &      36 &  30 &     1.2\\
 18.2  &     3  &      22 &  19 &     1.1\\
 21    &     3  &      4.5 &  8  &     0.6\\
\hline
\end{tabular}
}                                                                               \end{center}
\caption{ Fit of the FSS function of $\xi_{13}$ versus $\xi_{13}$. 
The symbols correspond to the
following lattice sizes: $(\circ)$ $ L\le 15$, $( \square )$ $ 15 < L \le 17$,
$(\triangle)$ $ 17 < L \le 22$, $(\lozenge)$ $ L > 22$. 
Dotted lines connect data at the same temperature. 
In the left frame only the data used in the fit are plotted,
while in the right frame all the data are showed.
The solid lines give the estimated error on
the fitted function (one standard deviation).
The graphs correspond to  $L_{\text{min}}=15.4$.}
\label{fig:fitfss_xi13}
\end{figure}

In Fig. \ref{fig:fssplot_xi13} we report the FSS function 
$F_\xi(z)$ for the correlation length $\xi_{13}$. The data show quite a 
good scaling, although at a closer look one may observe some 
small systematic deviations. 
The results of the fits are illustrated in Fig.~\ref{fig:fitfss_xi13}.
The figure is 
organized as follows. In the left frame  we plot the
residuals (the difference between the observed values and the fitting
function)  of the data used to perform the fit, i.e. the 
data points which fulfill
the condition $L \ge L_{\text{min},\xi}$; in the right frame the residual of
all the data points with respect to the same fit. This is useful to detect
possible systematic trends due to 
corrections to FSS that become smaller as the size of the system increases. 
The table under the frames reports 
the values of $\chi^2$  and $N_{\text{dof},\xi}$ for the fit as a
function of the cut $L_{\text{min},\xi}$ and of the number of 
parameters used $N_{{\rm par},\xi}$.

\begin{figure}
\vspace*{-2truecm} 
\begin{center}
\epsfig{file=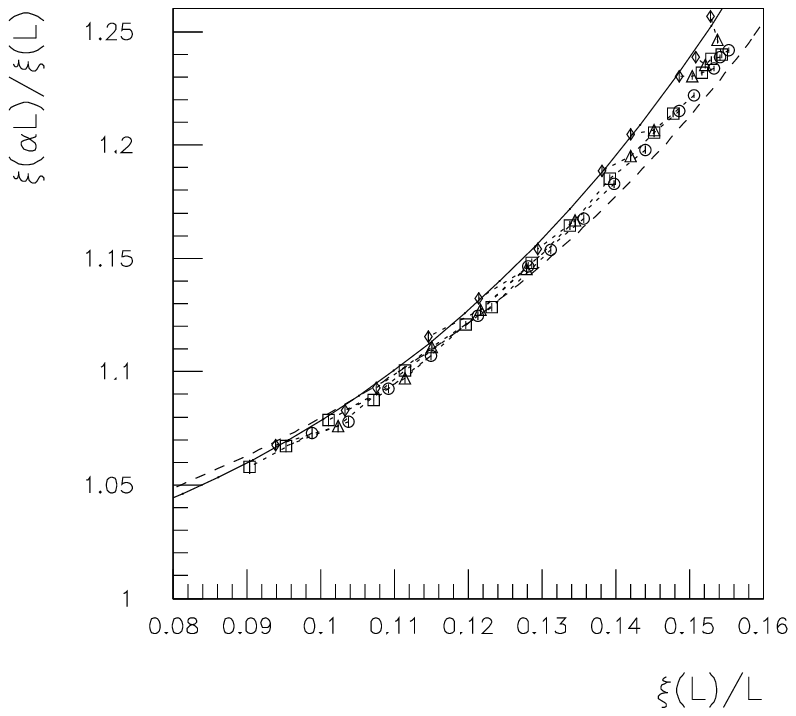,width=0.7\textwidth}
\end{center}
\vspace*{-1.5truecm} 
\caption{FSS plot for $\xi_{12}$ versus $\xi_{12}$. 
The symbols correspond to the
following lattice sizes: $(\circ)$ $ L\le 15$, $( \square )$ $ 15 < L \le 17$,
$(\triangle)$ $ 17 < L \le 22$, $(\lozenge)$ $ L > 22$.
Dotted lines connect data at the same temperatures.
The dashed line is the approximation \reff{eq:theoretical_pred},
the continuous line is the
result of the fit with $L_{\rm min} = 21$.}
\label{fig:fssplot_xi12}
\end{figure}

\begin{figure}
\begin{center}
\epsfig{file=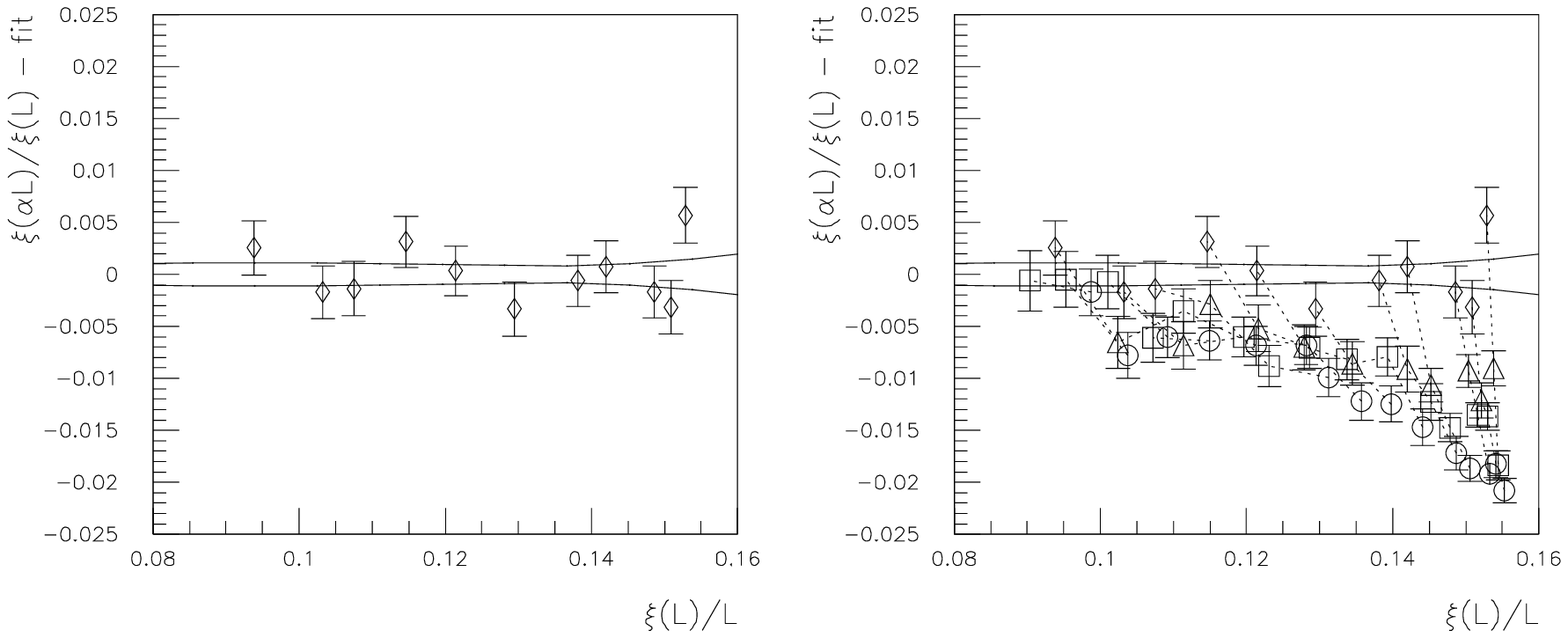,width=0.8\textwidth}
{\scriptsize
\begin{tabular}{|c|c|c|c|c|}
\hline
$L_{\text{min}}$ & $N_{\text{par}}$ & $\chi^2$ & $N_{\text{dof}}$ &
$\chi^2/N_{\text{dof}}$ \\ 
\hline
 14 &        3  &     277 & 49  &    5.6\\
 15.4 &       3   &    142  &34   &   4.2\\
 16.8 &       3   &    140  & 30  &    4.6\\
 18.2 &       3   &    83  & 19  &    4.4\\
 21 &       3   &    10  & 8   &    1.3\\
\hline
\end{tabular}
}
\end{center}
\caption{ Fit for the FSS function of $\xi_{12}$ versus $\xi_{12}$. 
The symbols correspond to the
following lattice sizes: $(\circ)$ $ L\le 15$, $( \square )$ $ 15 < L \le 17$,
$(\triangle)$ $ 17 < L \le 22$, $(\lozenge)$ $ L > 22$.
Dotted lines connect data at the same temperature.
In the left frame only the data used in the fit are plotted,
while in the right frame all the data are showed.
The solid lines give the estimated error on
the fitted function (one standard deviation).
The graphs correspond to  $L_{\text{min}}=21$.}
\label{fig:fitfss_xi12}
\end{figure}

We repeated the same analysis for $\xi_{12}$. As we discussed in Sec 2.2, 
this definition should not have a good behavior since it is not 
compatible with the symmetry of the low-temperature phase. 
This expectation is confirmed by the data, see
Figs.~\ref{fig:fssplot_xi12} and \ref{fig:fitfss_xi12}: 
The results show strong corrections to scaling 
and the fitted function clearly
underestimates the true $F_\xi(z)$. 

\begin{figure}
\vspace*{-2truecm} 
\begin{center}
\epsfig{file=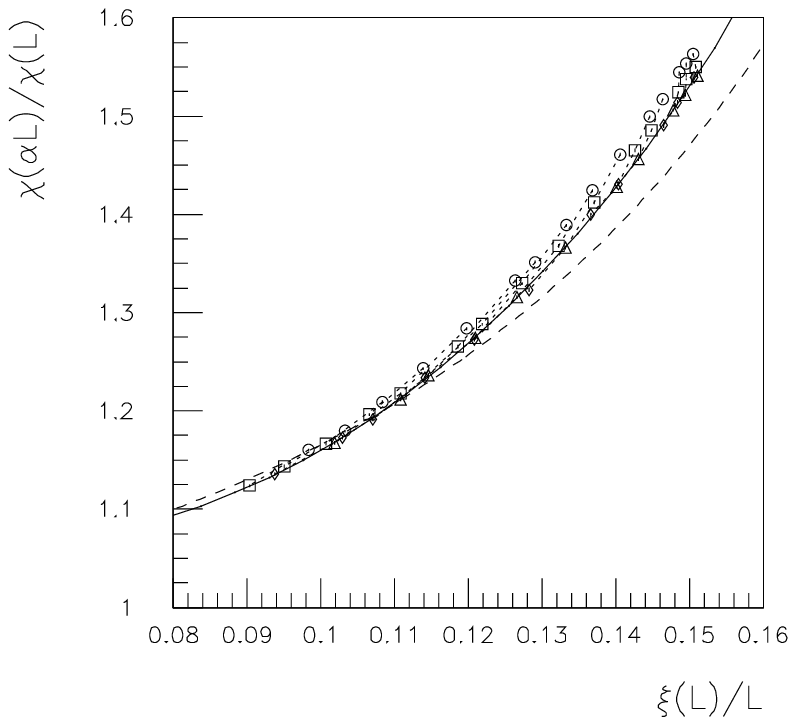,width=0.7\textwidth}
\end{center}
\vspace*{-1.5truecm} 
\caption{FSS plot for $\chi$ versus $\xi_{13}$. 
The symbols correspond to the
following lattice sizes: $(\circ)$ $ L\le 15$, $( \square )$ $ 15 < L \le 17$,
$(\triangle)$ $ 17 < L \le 22$, $(\lozenge)$ $ L > 22$.
Dotted lines connect data at the same temperatures.
The dashed line is the approximation \reff{eq:theoretical_pred_chi},
the continuous line is the
result of the fit with $L_{\rm min} = 18.2$.}
\label{fig:fssplot_s1_xi13}
\end{figure}

\begin{figure}
\begin{center}
\epsfig{file=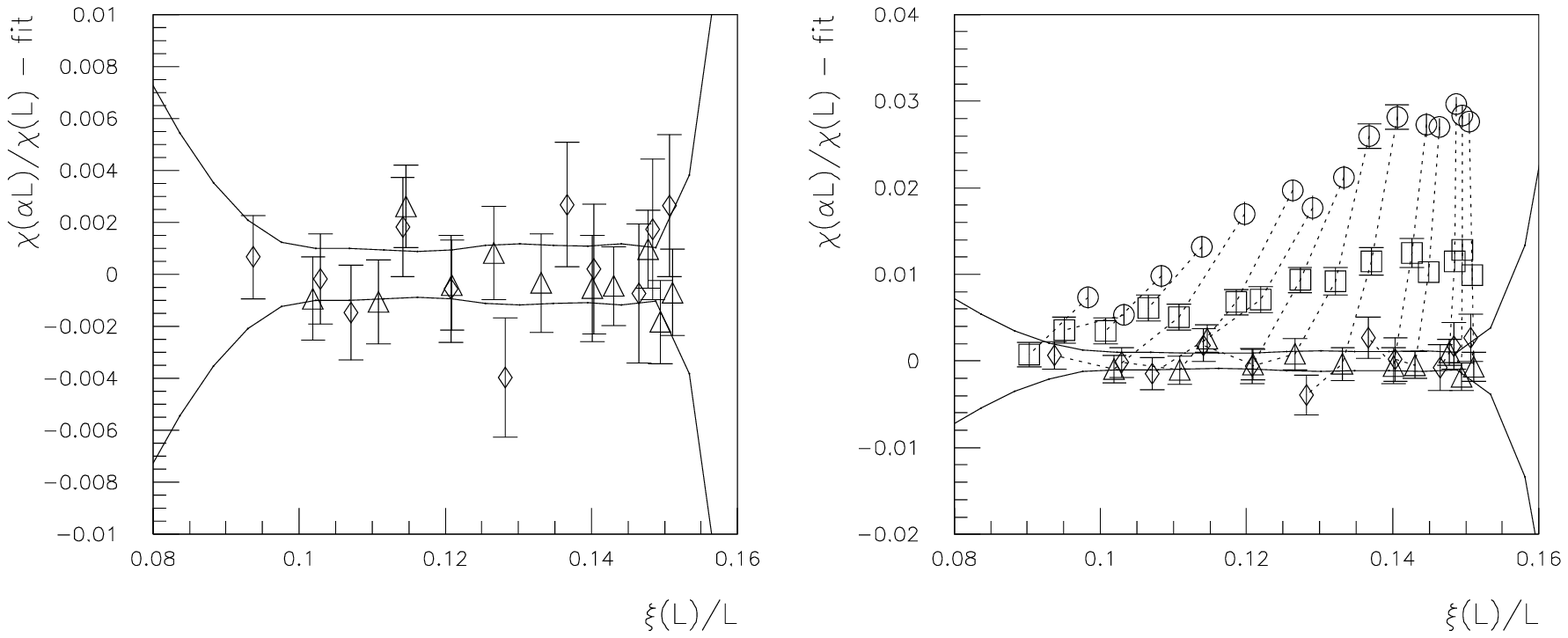,width=0.8\textwidth}
{\scriptsize
\begin{tabular}{|c|c|c|c|c|}
\hline
$L_{\text{min}}$ & $N_{\text{par}}$ & $\chi^2$ & $N_{\text{dof}}$ &
$\chi^2/N_{\text{dof}}$ \\
\hline
 14 &        3   &     2121 &  49 &     43\\
 15.4 &        3   &     224 &  34 &     6.6\\
 16.8 &        3   &     208 &  30 &     6.9\\
 18.2 &        5   &     13 &  17 &     0.78\\
 21 &        4   &     6.5 &  7  &     0.93\\
\hline
\end{tabular}
}
\end{center}
\caption{ Fit of the FSS function of $\chi$ versus $\xi_{13}$.
The symbols correspond to the
following lattice sizes: $(\circ)$ $ L\le 15$, $( \square )$ $ 15 < L \le 17$,
$(\triangle)$ $ 17 < L \le 22$, $(\lozenge)$ $ L > 22$.
Dotted lines connect data at the same temperature.
In the left frame only the data used in the fit are plotted,
while in the right frame all the data are showed.
The solid lines give the estimated error on
the fitted function (one standard deviation).
The graphs correspond to  
$L_{\text{min}}=18.2$.}
\label{fig:fitfss_s1_xi13}
\end{figure}

Then, we considered the susceptibility as a function of $\xi_{13}/L$. 
The FSS plot is reported in Fig. \ref{fig:fssplot_s1_xi13} 
and the  fit results in Fig. \ref{fig:fitfss_s1_xi13}. The corrections to scaling are 
larger than those for $\xi_{13}$, but it is quite nice to observe that 
the results for the largest lattices coincide perfectly. 

\begin{figure}
\epsfig{file=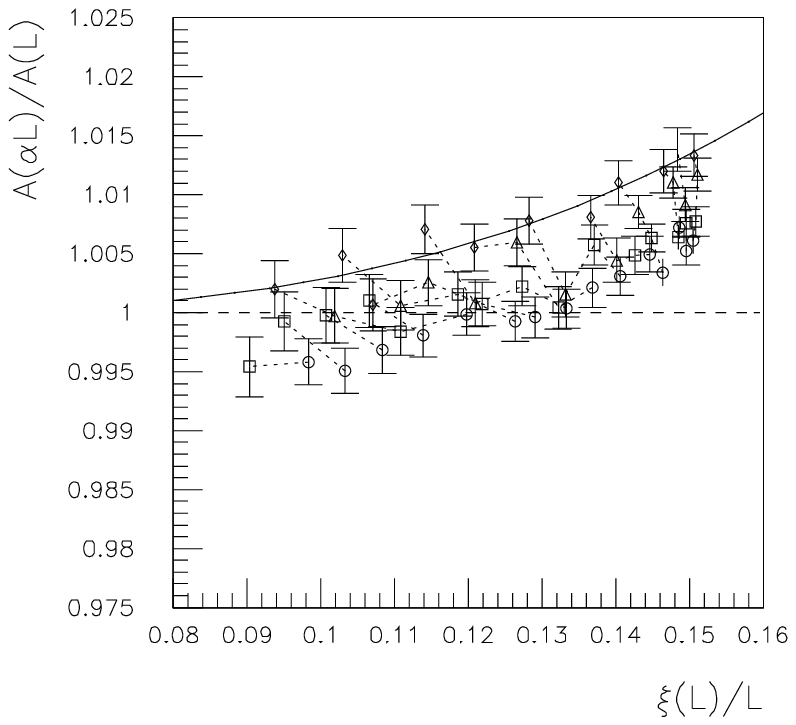,width=\textwidth}
\caption{FSS plot for $A_{13}$ versus $\xi_{13}$. 
The symbols correspond to the
following lattice sizes: $(\circ)$ $ L\le 15$, $( \square )$ $ 15 < L \le 17$,
$(\triangle)$ $ 17 < L \le 22$, $(\lozenge)$ $ L > 22$.
Dotted lines connect data at the same temperatures. 
The continuous line is the result of the fit with $L_{\rm min} = 21$.
}
\label{fig:fssplot_a13_xi13}
\end{figure}

As a further check we consider the ratio $A_{13}$ defined in Eq.
\reff{eq:amplitude}. Its FSS plot is reported in
Fig. \ref{fig:fssplot_a13_xi13}. We observe here clear corrections to 
scaling that make impossible the determination of $F_A(z)$.
Note that, if $\gamma/\nu =2$ as predicted by field theory, 
then the corrections must be nonmonotonic at least in the 
near-critical region. Indeed,
if $z^*$ is the critical-point constant defined in
Eq. \reff{def-zstar}---as we shall see $z^*\approx 0.15$---,
then $F_A(z^*) = 1$ if the exponents are the
field-theoretical ones.  On the contrary, our data seem to indicate
just the opposite: Apparently $F_A(z^*)$ is increasing with $L$ and
is larger than 1. 
Thus, we expect our FSS curves to be somewhat unreliable near 
$z\approx z^*$: This will be confirmed by a direct analysis that will
be performed in the next Section.

\begin{figure}
\vspace*{-2truecm} 
\begin{center}
\epsfig{file=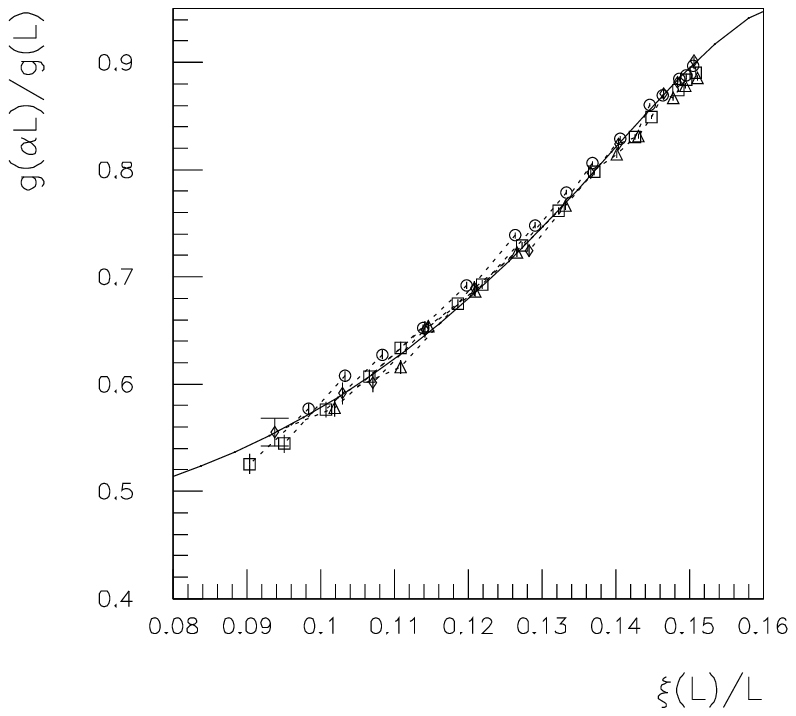,width=0.7\textwidth}
\end{center}
\vspace*{-1.5truecm} 
\caption{FSS plot for $g$ versus $\xi_{13}$. 
The symbols correspond to the
following lattice sizes: $(\circ)$ $ L\le 15$, $( \square )$ $ 15 < L \le 17$,
$(\triangle)$ $ 17 < L \le 22$, $(\lozenge)$ $ L > 22$.
Dotted lines connect data at the same temperatures.
The continuous line is the result of the fit with $L_{\rm min} = 21$.
}
\label{fig:fssplot_g_xi13}
\end{figure}

\begin{figure}
\begin{center}
\epsfig{file=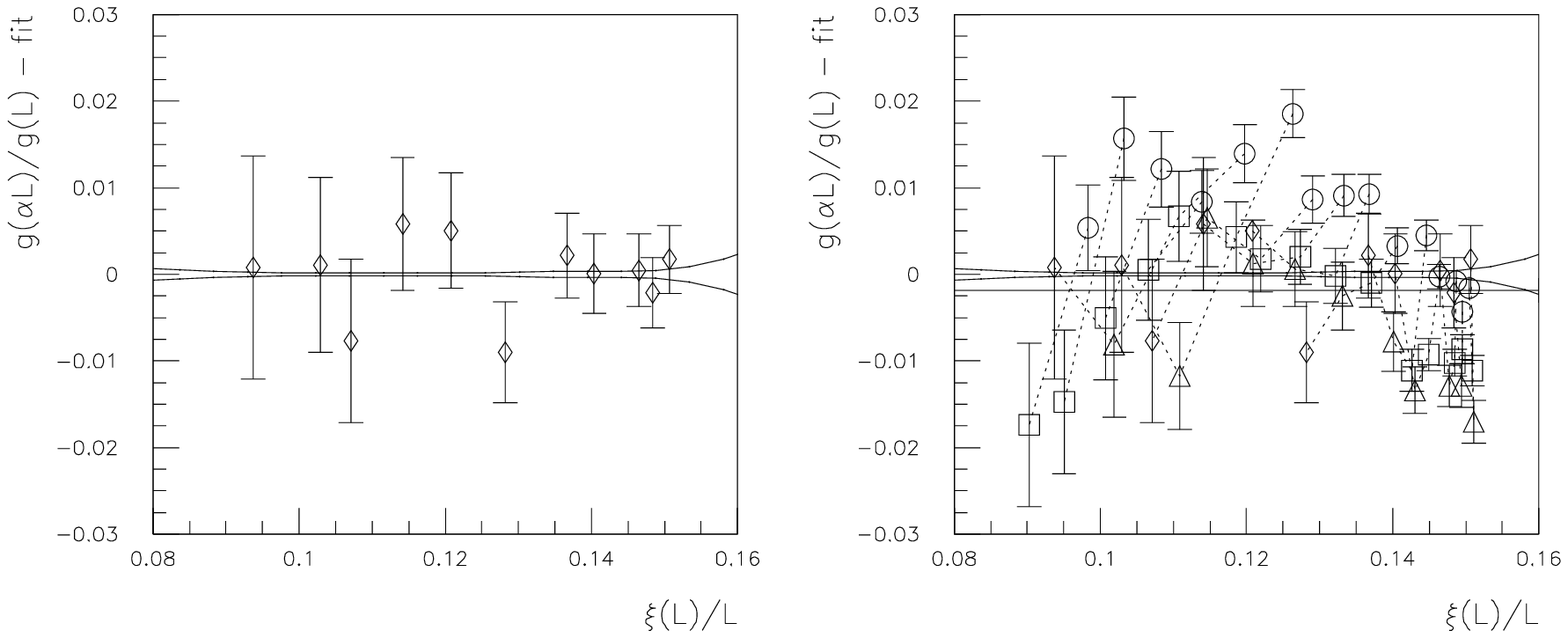,width=0.8\textwidth}
{\scriptsize
\begin{tabular}{|c|c|c|c|c|}
\hline
$L_{\text{min}}$ & $N_{\text{par}}$ & $\chi^2$ & $N_{\text{dof}}$ &
$\chi^2/N_{\text{dof}}$ \\
\hline
 14  &       3    &    260 &  49  &    5.3\\
 15.4  &       3    &    56 &  34  &    1.7\\
 16.8  &       3    &    56 &  30  &    1.9\\
 18.2  &       3    &    47 &  19  &    2.5\\
 21  &       4    &    4.7 &  7   &    0.68\\
 \hline
\end{tabular}
}
\end{center}
\caption{ Fit of the FSS function of $g$ versus $\xi_{13}$.
The symbols correspond to the
following lattice sizes: $(\circ)$ $ L\le 15$, $( \square )$ $ 15 < L \le 17$,
$(\triangle)$ $ 17 < L \le 22$, $(\lozenge)$ $ L > 22$.
Dotted lines connect data at the same temperature.
In the left frame only the data used in the fit are plotted,
while in the right frame all the data are showed.
The solid lines give the estimated error on
the fitted function (one standard deviation).
The graphs correspond to  $L_{\text{min}}=21$.}
\label{fig:fitfss_g_xi13} 
\end{figure}

\begin{figure}
\epsfig{file=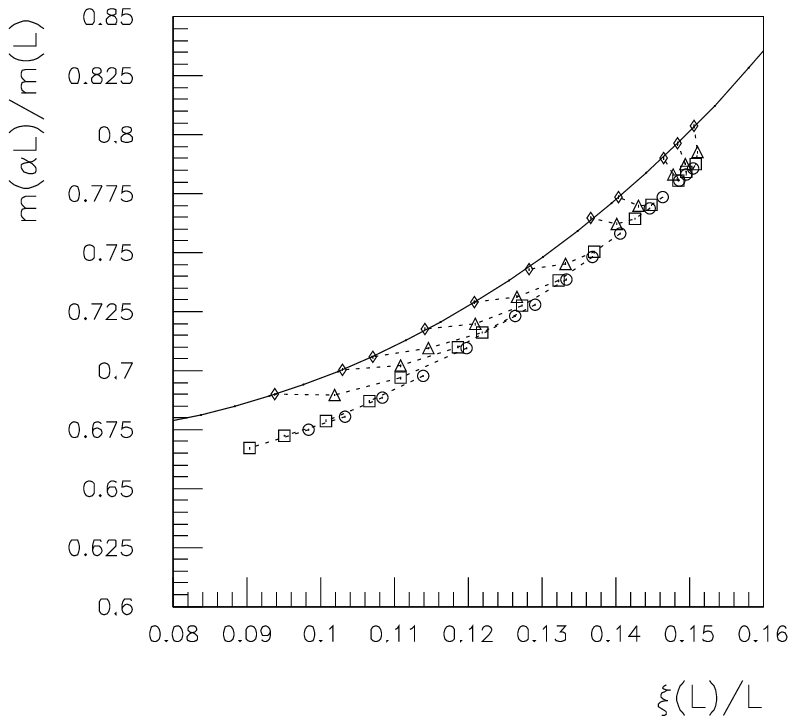,width=\textwidth}
\caption{FSS plot for $m$ versus $\xi_{13}$.
The symbols correspond to the
following lattice sizes: $(\circ)$ $ L\le 15$, $( \square )$ $ 15 < L \le 17$,
$(\triangle)$ $ 17 < L \le 22$, $(\lozenge)$ $ L > 22$.
Dotted lines connect data at the same temperatures.
The continuous line is the result of the fit with $L_{\rm min} = 21$.
}
\label{fig:fssplot_m_xi13}
\end{figure}

Finally, we report the FSS plots for the Binder parameter, 
Figs.~\ref{fig:fssplot_g_xi13}, \ref{fig:fitfss_g_xi13}, 
and for the magnetization,  
Fig.~\ref{fig:fssplot_m_xi13}.  The first one is quite good. There are 
tiny deviations, but they are not systematic as it can be seen 
from Fig. \ref{fig:fitfss_g_xi13}. The somewhat large $\chi^2$ found in the 
fits is probably an indication that the statistical errors are 
slightly underestimated. The FSS plot of $m$ is instead completely
unreliable. At the values of $L$ we consider, there are 
very large corrections to scaling and it is impossible to determine 
the FSS function $F_m(z)$. 

Let us now describe the extrapolation of $\xi$ and $\chi$ to the 
infinite-volume limit. 

With our preferred values of $L_{\text{min}}$ and
$N_{\text{par}}$ we obtain estimates of the FSS functions and then, 
for each triple $(L_{0},\xi_{0}(\beta),\chi_{0}(\beta)) =
(L,\xi(\beta,L),\chi(\beta,L))$ obtained from the simulations
we apply iteratively the map
\begin{equation}
\left(\begin{array}{c} 
 L_{n} \\ \xi_{n}(\beta) \\ \chi_{n}(\beta) 
\end{array}\right)
=
\left(
\begin{array}{c}
\alpha L_{n-1}\\\xi_{n-1}(\beta) \hat F_\xi(\xi_{n-1}(\beta)/L_{n-1})\\
\chi_{n-1}(\beta) \hat F_\chi(\xi_{n-1}/L_{n-1})\\ 
\end{array} \right) \; ,
\end{equation}
until the asymptotic limit $(\infty, \xi_\infty,\chi_\infty(\beta))$ is
reached. Thus, for each finite-volume MC result ${\cal O}(\beta,L)$, 
we obtain an extrapolated value ${\cal O}_\infty(\beta,L)$.

\begin{table}
\begin{center}
{
\scriptsize

\begin{tabular}{|c||c|c||c|c||c|c||c|c||}
\hline
 & 
\multicolumn{2}{|c|}{$L_{\text{min}} = 15.4$} &
\multicolumn{2}{|c|}{$L_{\text{min}} = 16.8$} &
\multicolumn{2}{|c|}{$L_{\text{min}} = 18.2$} &
\multicolumn{2}{|c|}{$L_{\text{min}} = 21$} \\
\hline
$\beta$ 
& $\chi_{\infty}$ & $R^2(\text{DOF})$ 
& $\chi_{\infty}$ & $R^2(\text{DOF})$ 
& $\chi_{\infty}$ & $R^2(\text{DOF})$ 
& $\chi_{\infty}$ & $R^2(\text{DOF})$ \\
\hline
0.27   & 12.049(42) & 0.8(2)  & 11.958(49) &   1.5(1) & 13.9(1.7) &   0.0(0)  &  & \\
0.275  & 13.878(51) & 3.2(2)  & 13.759(60) &   2.2(1) & 16.0(2.0) &   0.0(0)  &  & \\
0.28   & 16.352(64) & 0.8(2)  & 16.210(75) &   0.4(1) & 18.9(2.3) &   0.0(0)  &  & \\
0.285  & 19.714(82) & 5.1(2)  & 19.536(96) &   0.0(1) & 22.8(2.8) &   0.0(0)  &  &  \\
0.29   & 24.683(87) & 7.9(5)  & 24.49(10) &   5.0(4)  & 29.1(3.4) &   1.3(3)  &  26.03(79) &   1.3(2)\\
0.295  & 32.77(13)  & 9.3(5)  & 32.44(15) &   5.6(4)  & 36.4(4.6) &   3.0(3)   & 34.0(1.2) &   0.8(2)\\
0.297  & 37.49(16)  & 15.5(5) & 37.10(18) &   11.1(4) & 38.6(5.1) & 9.3(3) & 38.8(1.3) &   8.7(2)\\
0.3    & 48.01(21) &  5.5(5)  & 47.51(26) &   5.8(4)  & 51.8(6.6) &   4.0(3)   & 49.4(1.6) &   3.7(2)\\
0.3025 & 62.06(28) &  3.6(5)  & 61.47(33) &   3.6(4)  & 70.0(8.7) &   1.6(3)   & 64.3(2.1) &   1.5(2)\\
0.305  & 87.02(42) &  14.0(5) & 86.35(48) &   8.4(4)  & 91(12) &   4.4(3)   & 89.8(3.0) &   0.8(2)\\
0.3075 & 146.30(83) & 4.8(5)  & 145.01(95) &   4.9(4) & 158(20) &   3.4(3)  & 150.2(5.1) &   0.1(2)\\
0.3085 & 201.4(1.4) & 11.2(5) & 200.0(1.5) &   7.9(4) & 211(27) &   5.7(3)     & 205.8(7.1) &   1.2(2)\\
0.31   & 474.5(7.3) & 9.4(5)  & 472.6(7.7) &   5.4(4) & 504(65) &   4.2(3)    & 478(19) &   1.6(2)\\
0.3105 & 821(24) &    11.8(5) & 813(24)    &   7.8(4) & 829(110)  & 6.4(3)       & 823(37) &   0.4(2)\\
\hline
\end{tabular}
}
\end{center}
\caption{Infinite-volume results for $\chi$, obtained from the 
extrapolation of finite-size data, for different values of 
  $L>L_{\text{min}}$. $R^2$ is the $\chi^2$ defined in Eq.
  \reff{R2}, and DOF is the number of degrees of freedom.}
\label{table:extr_chi}
\end{table}

To combine the estimates ${\cal O}_\infty(\beta,L_i)$ with the 
same value of $\beta$ and
have a consistent estimate $\obs_\infty(\beta)$, we proceed as in
Refs. \cite{Caracciolo95,Caracciolo95b}.
We generate fake data sets, which are obtained by replacing each MC result
$\cal O$ with error $\Delta \cal O$ by ${\cal O} + r \Delta \cal O$, where 
$r$ is a normally distributed variable with zero mean and unit variance. 
Then, for each fake data set, 
using the same $L_{{\rm min},\obs}$ and $N_{\rm par,\obs}$, 
we compute the FSS curves and perform the extrapolations. Thus, for each 
$\beta$ and $L$ that we consider in the simulation we obtain 
a sequence of extrapolated values $\obs_k^\infty(\beta,L)$, 
$k=1,\ldots,M$---typically $M \approx 10^3$---where 
$\beta$ and $L$ indicate that $\obs_k^\infty(\beta,L)$ is obtained by 
extrapolating 
the (fake) finite-volume result $\obs_k(\beta,L)$.
We use these results to compute the covariance of the estimates 
at fixed $\beta$
\be 
C_{ij}(\beta) = {1\over M} \sum_{k=1}^M \left({\cal O}_k^\infty(\beta,L_i) - 
                 {\cal O}_\infty(\beta,L_i)\right)
                \left({\cal O}_k^\infty(\beta,L_j) -
                 {\cal O}_\infty(\beta,L_j)\right).
\end{equation}
Then, we  consider the weighted average
\be
\widehat{\obs}(\beta) = 
{\sum_{i,j} (C^{-1}(\beta))_{ij} {\cal O}_\infty(\beta,L_j)\over 
      \sum_{i,j} (C^{-1}(\beta))_{ij}} ,
\end{equation}  
which minimizes the residual sum of squares
\be
R^2(\beta) = \sum_{i,j} (C^{-1}(\beta))_{ij} 
       ({\cal O}_\infty(\beta,L_i) - \widehat{\obs}(\beta)) 
       ({\cal O}_\infty(\beta,L_j) - \widehat{\obs}(\beta)) .
\label{R2}
\end{equation}
If the variables ${\cal O}_\infty(\beta,L_i)$, $i=1,\ldots,n(\beta)$, 
are Gaussian with covariance 
$C(\beta)$, then $R^2(\beta)$ is distributed as a
$\chi^2$ random variable of $n(\beta)-1$ degrees of freedom (DOF). Then, the
deviations of $R^2(\beta)$ from $n(\beta)-1$ 
are a measure of the consistency of the
different extrapolations and $\widehat{\obs}(\beta)$ is our best estimate of
$\obs_\infty(\beta)$. The estimated error is  
\be
\Delta \widehat{\obs}(\beta) = 
     \left(\sum_{i,j} (C^{-1}(\beta))_{ij}\right)^{-1/2}.
\end{equation}     
The parameters of our preferred extrapolations are
$L_{\text{min}}=15.4$ and $N_{\text{par}}=3$ for $\xi_{13}$
and $L_{\text{min}}=18.21$ and $N_{\text{par}}=5$ for $\chi$. 
The corresponding FSS functions are given by
\begin{eqnarray}
F_\xi(z) &=& 
1 - 2.14297\,z^2 - 171.098\,z^3 + 535.11\,z^4,  \\
F_\chi(z) &=& 
1 - 113.951\,z^2 + 3053.15\,z^3 - 38192.7\,z^4 + 204378\,z^5 -
394893\,z^6\; .
\end{eqnarray}
The results are reported in 
Tables~\ref{table:extr_xi} and \ref{table:extr_chi}.
All extrapolations are consistent, i.e. 
$R^2(\beta) \approx n(\beta) - 1$,  except in a few
isolated cases (for example the extrapolation of $\xi_{13}$ 
at $\beta=0.3105$ for $L_{\text{min}}=16.8$). Note also that
the direct extrapolation 
of $A_{13}$---here we use the larger lattices $L\ge 21$ to compute 
$F_A(z)$---is in good agreement with the results obtained from the 
extrapolated estimates of $\chi$ and $\xi$, in spite of the fact 
that we do not have a reliable determination of $F_A(z)$. 
Clearly the errors due the neglected corrections to scaling
are small compared to the extrapolation errors.

\begin{figure}
\begin{center}
\epsfig{file=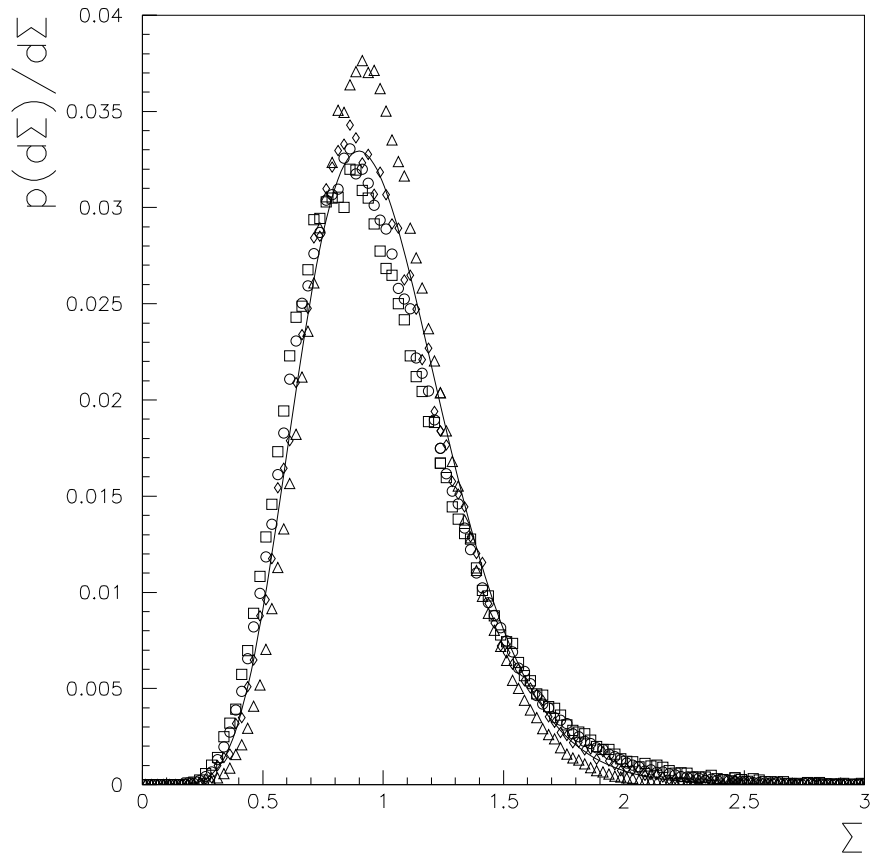,width=0.8\textwidth}
\end{center}
\caption{Distribution function of the variable $\Sigma$. 
  The symbols refer to different 
  values of $\beta$ and $L$: $\circ$ $(\beta=0.27,L=16)$,
  $\square$ $(\beta=0.29,L=28)$, $\triangle$ $(\beta=0.311,L=16)$, $\lozenge$
  $(\beta=0.311,L=28)$. The solid line is the theoretical prediction as
  discussed in the text. }
\label{Mag-distribution}
\end{figure}

We wish finally to check the prediction 
\reff{prediction-M2-distribution} for the probability distribution of 
$M^2$. This cannot be done straightforwardly since we did not 
save the values of $\phi({\bf k}_{0,1})$ for each 
single configuration, but rather the sum of the values of 
$\phi({\bf k}_{0,1})$ for $N=10$ independent configurations---the 
ten configurations that were simulated together by our 
multispin program---. This means that we cannot determine from our
data the distribution of $M^2$, but rather that of 
\be 
       \Sigma = {1\over N} \sum_{i=1}^N \sigma_i
\ee
where 
\be
\sigma_i = {M^2 \pa{L} \pe{L}\over \chi},
\ee
where the sum is over $N$ independent systems. Since the variables 
$\sigma_i$ are squares of normalized
complex Gaussian random variables (see Eq.~\reff{prediction-M2-distribution}), 
$\Sigma$ is distributed as a
reduced $\chi^2$ variable with $2 N$ degrees of freedom (the two is due to the 
fact that the random variables  are complex).
In Figure~\ref{Mag-distribution} we report the distribution of
$\Sigma$ for some values of $\beta$ and $L$, together with the 
theoretical prediction, that does not include any free parameter.
As we can see the collapse is fairly good 
especially for the largest lattice near the critical temperature.

\subsection{Determination of the critical indices} \label{sec.5-3}

In order to determine the critical exponents we performed two different
analyses. First, we considered the infinite-volume estimates of
$\xi_\infty(\beta)$ and $\chi_\infty(\beta)$ and performed a fit to 
a power law in the 
reduced temperature $t = 1-\beta/\beta_c$. Then, we considered the FSS
behavior determining the exponents from Eqs. \reff{eq:critical_scaling}
and \reff{eq:dummy_dimensions}.

Let us begin with the infinite-volume correlation length 
$\xi_{13}$ (in the following we 
simply write it as $\xi$). We performed several fits of the 
form
\be
\label{fit1-xi}
\xi_\infty(\beta) = B_\xi t^{-\nu}, 
\end{equation}
using in each case only the estimates of $\xi_\infty(\beta) $
with $\beta\ge \beta_{\rm min}$. 
We find the results reported in Table \ref{results-nu-fit1}.
\begin{table}
\begin{center}
\begin{tabular}{|l|l|l|l|l|l|}
\hline
$\beta_{\rm min}$ & $\nu$ & $B_\xi$ & $\beta_c$ & $\chi^2$/DOF & DOF \\
\hline
0.270  & 0.556(15) & 0.552(22)& 0.31135(15) & 6.1 & 11 \\
0.280  & 0.542(13) &0.577(21)&  0.31126(11) &2.8 &9 \\
0.290  & 0.531(11) & 0.600(21)& 0.311207(80) &1.3 &7 \\
0.300  & 0.518(15)&0.629(32)&   0.311153(73) &0.53 &4 \\
0.305  & 0.522(56) &0.62(14) &  0.31116(21) &0.91 &2 \\
0.3075 & 0.54(41) & 0.56(93) &  0.3112(13)& 1.0 &1 \\
\hline
\end{tabular}
\end{center}
\caption{Results for the fit \protect\reff{fit1-xi} for several values of 
$\beta_{\rm min}$. 
}
\label{results-nu-fit1}
\end{table}  
One clearly observes the presence of corrections to scaling:
The estimate of $\nu$ decreases as 
$\beta_{\rm min}$ increases until the fit looses predictivity. 
We may try to take the corrections into account, 
by performing a fit of the form
\be
\label{fit2-xi}
\xi_\infty(\beta) = B_\xi t^{-\nu} (1 + C_\xi t^\theta). 
\end{equation}
Since no theoretical estimate is available for the 
exponent $\theta$ and our data are not precise enough to keep 
$\theta$ as a free parameter,  we have performed several 
fits, using ``reasonable" values of $\theta$. 

If we assume that the leading correction is the analytic one, i.e. $\theta=1$,
the fit improves substantially. The smallest value of $\chi^2/\text{DOF}$ is attained for 
the lowest possible value of $\beta_{\rm min}$, i.e. $\beta_{\rm min} = 0.27$. 
Then, we obtain 
\begin{eqnarray} 
    \nu &=& 0.4970(93), \nonumber \\
    \beta_c &=& 0.311109(44), \nonumber \\
    B_\xi &=&  0.693(24), \nonumber \\
    C_\xi &=&  -0.90(13),
\end{eqnarray}
with $(\chi^2/\text{DOF})=0.366$ and $\text{DOF} = 10$.

If we consider $\theta = 0.5$, we obtain again the best fit for $\beta_{\rm
  min} = 0.27$, yielding
\begin{eqnarray}
    \nu &=& 0.459 (16), \nonumber  \\
    \beta_c &=& 0.311058(58) \nonumber \\
    B_\xi &=&  0.866(63)  \nonumber \\
    C_\xi &=& -0.658(92) 
\end{eqnarray}                                                                  
with $(\chi^2/\text{DOF})=0.526$ and $\text{DOF}=10$. Note that 
the $(\chi^2/\text{DOF})$ is here slightly larger that in the case 
$\theta=1$, providing some (very weak) evidence in favor of an 
exponent $\theta=1$.

These results indicate clearly that $\nu\ltapprox 1/2$, with $\nu$ 
decreasing with decreasing $\theta$. If $\theta=1$, the result is perfectly 
compatible with the theoretical prediction \reff{sec4:infvol}.
We also tried to see the effect of an additional analytic correction.
If we consider 
\be
\xi_\infty(\beta) = B_\xi t^{-\nu}\left( 1+C_\xi t+D_\xi t^2 \right ), 
\end{equation}
we obtain for $\beta_{\text{min}} = 0.27$
\begin{eqnarray}
\nu &=&  0.501(24),  \nonumber   \\
\beta_c &=& 0.311118(73),  \nonumber \\
B_\xi &=&0.681(72), \nonumber \\
C_\xi &=& -0.75(86), \nonumber \\
D_\xi &=& -0.6(3.8),
\end{eqnarray}
with $(\chi^2/\text{DOF})=0.399$ and $\text{DOF}=9$. As it can be seen 
from the errors on $C_\xi$ and $D_\xi$, the addition of another 
fitting parameter is not justified. Our data are not precise enough 
to allow the determination of two correction-to-scaling terms.

Finally, assuming the theoretical prediction $\nu = 1/2$, we compute 
our best estimate of the leading correction-to-scaling exponent. We 
redo the fit \reff{fit2-xi}, fixing $\nu = 1/2$ and keeping
$\theta$ as a free parameter. Again, the best fit is obtained 
by using all data, i.e. with $\beta_{\rm min} = 0.27$. We obtain 
$\theta = 1.09(22)$, $\beta_c = 0.311114(28)$ with 
$(\chi^2/\text{DOF})=0.356$ and $\text{DOF}=10$. Since an analytic 
correction should be present, and therefore $\theta\le 1$, this 
result confirms the analytic nature of the first correction term.

A further refinement can be obtained if we assume  the theoretical 
prediction $\nu = 1/2$ and $\theta=1$. This provides our best estimate 
of the critical temperature. If we perform the fit \reff{fit2-xi} 
fixing $\nu$ and $\theta$, we obtain
\begin{eqnarray}
\beta_c &=& 0.311122(20),  \nonumber \\
B_\xi &=&   0.6854(30), \nonumber \\
C_\xi &=& -0.860(46), 
\label{betac-constrained-nu-theta}
\end{eqnarray}
with $(\chi^2/\text{DOF})=0.35$ and $\text{DOF}=11$.

We can repeat the same analysis for the susceptibility $\chi$. We perform a fit
of the form 
\be
\chi_\infty(\beta) = B_\chi t^{-\gamma} ,
\end{equation}
and obtain for $\beta_{\rm min} = 0.27$
\begin{eqnarray}
\gamma &  =  &  0.995(34), \nonumber  \\
\beta_c &  = &  0.31118(11), \nonumber  \\
B_\chi   & =  & 1.90(19), 
\end{eqnarray}
with $(\chi^2/\text{DOF}) = 0.076$ and $\text{DOF}=11$. 
The $\chi^2$ is very small, indicating that probably the errors on 
$\chi_\infty(\beta)$
are overestimated. The final result for $\gamma$ is perfectly compatible
with the theoretical estimate \reff{eq:theoretical_pred_chi}. Also, the 
estimate of $\beta_c$ is in good agreement with the estimates of $\beta_c$ 
obtained from the analysis of the correlation length. Although the 
goodness of the fit does not require it, we also performed a fit including 
an analytic correction. For $\beta_{\rm min} = 0.27$, we obtain 
$\gamma = 0.98(10)$, $\beta_c = 0.31116(20)$, 
$(\chi^2/\text{DOF}) = 0.0825$ and $\text{DOF}=10$. Clearly nothing changes.
If we fix $\gamma = 1$, we obtain a more precise estimate of 
$\beta_c$. For $\beta_{\text{min}}=0.27$ we have
\begin{eqnarray} 
\beta_c &=& 0.311198(51),  \nonumber \\
B_\chi &=&   1.877(43), 
\label{betac-constrained-gamma}
\end{eqnarray}
with $(\chi^2/\text{DOF})=0.070$ and $\text{DOF}=12$. 
This estimate of $\beta_c$ is in good agreement with the result
\reff{betac-constrained-nu-theta} for the correlation length.

We have also performed a fit of the form 
\begin{equation}
\log \chi(\beta) = a_1 \log \xi(\beta) + b_1  ,
\label{eq:fit-logchi-logxi}
\end{equation} 
that allows to determine 
$a_1 = \gamma/\nu$ independently of the critical
temperature.  The results are reported in Table~\ref{results-fit-logchi-logxi}.
\begin{table}
\begin{center}
\begin{tabular}{|l|l|l|l|l|l|}
\hline
$\beta_{\rm min}$ & $a_1$ & $b_1$ & $\chi^2$/DOF & DOF \\
\hline
 0.270   & 1.856(35) & 1.652(51)   &0.096  & 12 \\
 0.280   & 1.867(42)  & 1.631(66)  & 0.10 & 10 \\
 0.290   & 1.883(52)  & 1.600(88)   & 0.10 &8 \\
 0.300   & 1.899(51) & 1.57(10) & 0.041 &5  \\
 0.305  & 1.921(72)  & 1.52(15) & 0.026 &3 \\
 0.3075 & 1.90(11)  & 1.58(24)  & 0.018 &2 \\
\hline
\end{tabular}
\end{center}
\caption{Results for the fit \protect\reff{eq:fit-logchi-logxi} for several values of 
$\beta_{\rm min}$. 
}
\label{results-fit-logchi-logxi}
\end{table}  
There are clear corrections to scaling, with the estimates of $a_1$ 
increasing towards 2 as $\beta_{\rm min}$ increases. These results 
are perfectly consistent with what we observed before. If no 
correction to scaling is included, the effective $\nu$ varies between 
0.56 and 0.52 so that one would expect $\gamma/\nu$ to vary between
1.8 and 1.9, which is exactly what we find here.

In conclusion, we find good evidence that $\xi$ and $\chi$ scale in the 
infinite-volume limit as predicted by field theory. 
If $\gamma = 1$ and $\nu = 1/2$, we do not find evidence of 
nonanalytic corrections with  $\theta< 1$. Assuming the 
theoretical predictions for the exponents we obtain for the critical 
temperature
\begin{eqnarray}
 \beta_c & = 0.31115(10), \label{betacfinale} 
\end{eqnarray}
where we have used the results
\reff{betac-constrained-nu-theta} and \reff{betac-constrained-gamma}. 
The given error is a conservative estimate that should take into account 
both statistical and systematic uncertainties.
The result for $\beta_c$ should be compared with the existing determinations:
\begin{equation}
\beta_c = \cases{0.3108(11) & $\hskip 1truecm$ Ref. \protect\cite{Leung91b}; \cr
                 0.3125(13) & $\hskip 1truecm$ Ref. \protect\cite{Wang96}. \cr
            }
\end{equation}
Our result \reff{betacfinale} is in good agreement with both estimates,
although much more precise.

The critical exponents can also be computed from the FSS scaling curves.
First, we determine $z^*$ by solving the equation 
$\alpha = F_\xi(z^*)$ and by using our determination of $F_\xi(z)$. 
We find $z^* = 0.1510(1)$. Then, for each observable, we compute 
$w^* = F_\obs(z^*)$ and $\gamma_\obs/\nu$, 
see Eq. \reff{eq:dummy_dimensions}. The results are reported in 
Table \ref{table-exponents-FSS}. For the magnetization $m$, cf.
Eq. \reff{magnetization},  we extrapolated the data with $L\ge 21$, 
although they are probably not asymptotic. 
Thus, the correct estimate of $\beta/\nu$  could slightly decrease.
\begin{table}
\begin{center}
\begin{tabular}{|l|l|l|}
\hline 
 & $w^*$ & $\gamma_\obs/\nu$ \\
\hline
$\chi$ & 1.543(2) & 1.944(5)  \\
$A_{13}$ & 1.013(1)  &  0.060(5) \\   
$g$ &  0.902(4) & $-$0.46(2) \\
$m$ &  0.804(7) & $-$0.98(4) \\
\hline
\end{tabular}
\end{center}
\caption{Estimates of $\gamma_\obs/\nu$ for several different observables.}
\label{table-exponents-FSS}
\end{table}

The estimate of $\gamma/\nu$ is not far from the expected value $\gamma/\nu=2$.
The result is slightly lower, which may indicate that the curve varies
more strongly than predicted by our naive extrapolation of the FSS curve. 
The reason of the difference clearly appears if we consider $A_{13}$ 
whose scaling dimension is $2-\gamma/\nu$. 
Here, we find a slightly positive value
related to the already observed fact that $F_A(z^*)\not=1$. 
Clearly for $z\approx z^*$ there are corrections to scaling that 
we have not taken into account. A better analysis will be presented below.

\begin{figure}
\vspace*{-1.5truecm} 
\begin{center}
\epsfig{file=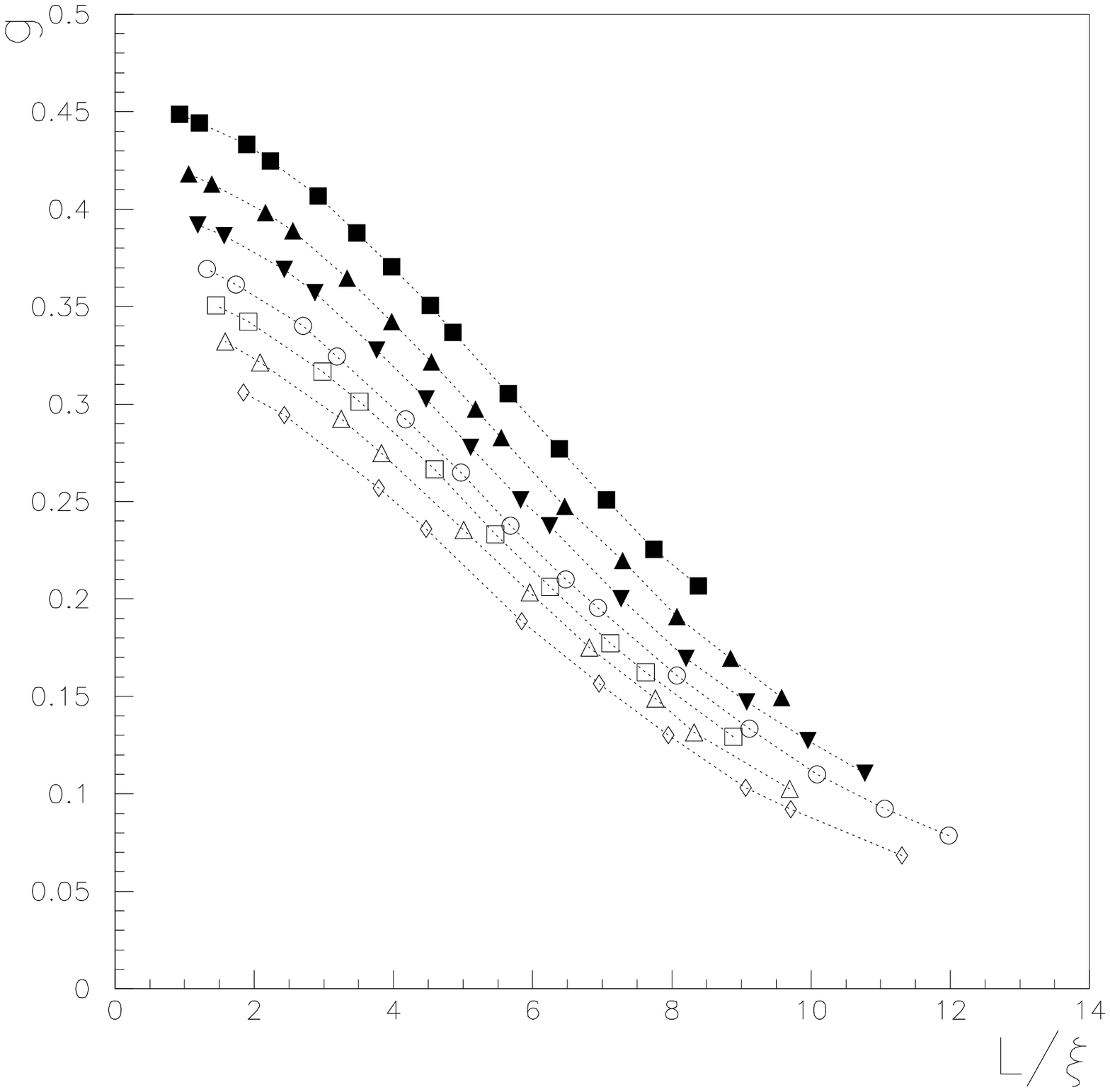,width=0.6\textwidth}
\end{center}
\vspace*{0.truecm} 
\caption{Binder cumulant: $g$ versus $\xi_{13}/L$. 
The symbols correspond
to the following lattice sizes: $L$=
$14(\blacksquare)$, $16 (\blacktriangle)$, $18 (\blacktriangledown)$,
$20 (\circ  )$, $22 (\square) $, $24 (\triangle)$, $28 (\lozenge)$. 
}
\label{Binder-FFSplot-I}
\end{figure}

\begin{figure}
\vspace*{-1.5truecm}
\begin{center}
\epsfig{file=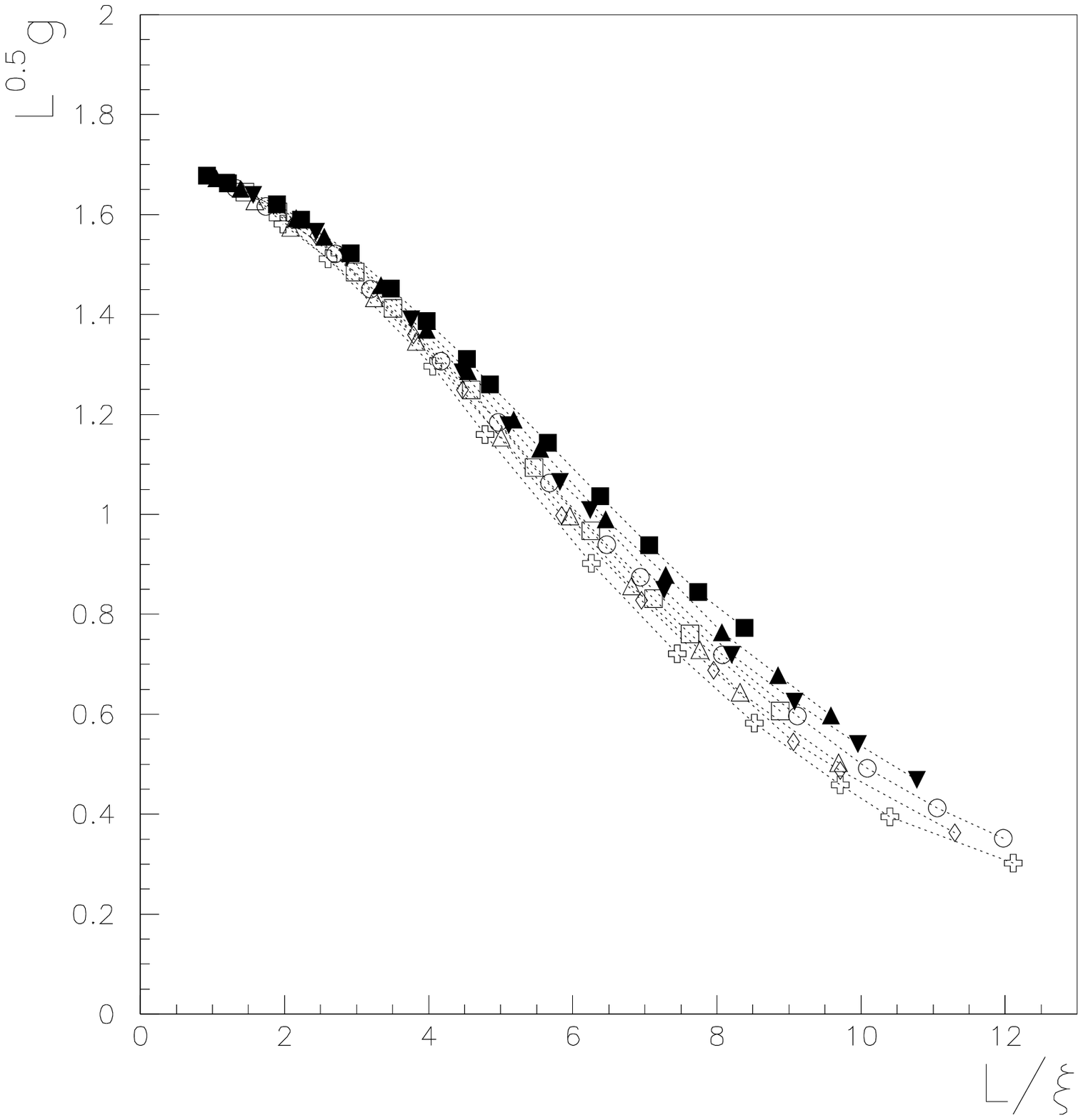,width=0.6\textwidth}
\end{center}
\vspace*{0.truecm} 
\caption{Rescaled Binder cumulant: $L^{0.5} g$ versus $\xi_{13}/L$. 
The symbols correspond 
to the following lattice sizes: $L$=
$14(\blacksquare)$, $16 (\blacktriangle)$, $18 (\blacktriangledown)$,
$20 (\circ  )$, $22 (\square) $, $24 (\triangle)$, $28 (\lozenge)$.  
}
\label{Binder-FFSplot-II}
\end{figure}

Although our estimate of the FSS curve for $m$ is quite bad, the result
for $\beta/\nu$ is in very good agreement with the theoretical 
estimate $\beta/\nu=1$. Finally, let us consider the Binder parameter 
that apparently vanishes at the critical point as $L^{-0.46}$. 
The systematic decrease of the renormalized coupling 
can be seen more clearly in Fig. \ref{Binder-FFSplot-I}. The fact that
$g$ scales with a negative exponent can also be obtained 
with a scaling plot, by looking for $q$ so that
\be 
g L^q \approx F(\xi/L).
\end{equation}
The optimal $q$ is $q \approx 0.5$, in agreement with the 
result of Table \ref{table-exponents-FSS}. The corresponding plot 
is given in Fig. \ref{Binder-FFSplot-II}.

It is also possible to compute the critical exponents by using the 
raw data near the critical point. The idea is to use Eq. 
(\ref{eq:critical_scaling}), keeping into account the fact that 
our data are not exactly at $\beta_c$. In this case 
we should add a correction term and use
\be
{\cal O}(\beta,L) = L^{\gamma/\nu} 
    \left(1 + a (\beta - \beta_c) L^{1/ \nu} + \ldots\right),
\label{FSS-quasiatTc}
\ee
where we have neglected corrections to scaling. 
Equation \reff{FSS-quasiatTc} is valid in the limit in which 
$(\beta - \beta_c) L^{1/\nu}$ is a small correction.

We have considered our data at $\beta = 0.311$ that are quite near to the
critical point and we have performed two sets of fits. In the 
first fit, we have neglected the correction term
in Eq. \reff{FSS-quasiatTc}, by considering
\begin{equation}
  \xi(\beta,L) = Z_\xi L^{\rho}, \qquad
  \chi( \beta,L) = Z_\chi L^{\gamma/\nu}.
  \label{eq:critical_scaling_2}
\end{equation}
Using our data for 
$L=14,16,18,20,22,24,28$, we have performed a series of fits,
including each time
only the data with $L \ge L_{\rm min}$. For $\xi$ the results are stable 
with $L_{\rm min}$. The smallest $\chi^2$ is obtained for $L_{\rm min}=18$.
Correspondingly, 
\begin{table}
\begin{center}
\begin{tabular}{|l|l|l|l|l|l|}
\hline
$L_{\rm min}$ & $\gamma/\nu$ & $Z_\chi$ & $\chi^2$/DOF & DOF \\
\hline
14 & 1.975(24) & 0.0902(63) & 62 &5 \\
16 & 1.955(24) & 0.0958(67) & 23  & 4 \\
18 & 1.939(27) & 0.1004(77) & 10  & 3 \\
20 & 1.925(31) & 0.105(10) & 3.207 & 2 \\
22 & 1.94(17)  & 0.101(54) & 3.204 & 1 \\
\hline
\end{tabular}
\end{center}
\caption{Results for the fit~(\ref{eq:critical_scaling_2}) 
for several values of $L_{\rm min}$. 
}
\label{results-fit-critical-scaling}
\end{table}  
\be
\rho = 0.9923(85), \qquad
Z_\xi = 0.1544 (40),
\end{equation}
with $\chi^2/\text{DOF}=1.510$ and $\text{DOF}=3$. This is in perfect 
agreement with the expected value $\rho = 1$ and confirms the 
correct behavior of the correlation length. The 
analysis of the susceptibility gives the results 
reported in Table~\ref{results-fit-critical-scaling}. They show a clear 
downward trend and seem to indicate that $\gamma/\nu < 2$.
This conclusion is however incorrect. The trend is due to the fact 
that we have not taken into account the fact that we are not exactly
at the critical point. The correction term in Eq. \reff{FSS-quasiatTc}
plays an important role, as we now show.
We repeat the fit by using, in accordance with 
Eq. \reff{FSS-quasiatTc}, 
\begin{equation}
\chi(\beta,L) = 
Z_\chi L^{\gamma/\nu} \left(1 + X_\chi L^{2}\right),
\label{eq:critical_scaling_3}
\end{equation}
where we have set $\nu = 1/2$ in the correction term. 
The results are reported in Table~\ref{results-fit-critical-scaling-2}.
\begin{table}
\begin{center}
\begin{tabular}{|l|l|l|l|l|l|}
\hline
$L_{\rm min}$ & $\gamma/\nu$ & $Z_\chi$ &$X_\chi$ & $\chi^2$/DOF & DOF \\
\hline
14 & 2.060(43) &0.0736(77)  &  $-$0.000120(56)&9.3  &4 \\
16 & 2.031(80)  & 0.079(16)   &$-$0.000091(90) & 7.5  & 3 \\
18 & 2.00(21)  &0.086(48)  &   $-$0.00006(21) & 8.7  &2 \\
20 & 1.9(1.1)  & 0.11(34)   &0.0000(11) & 5.78 &1 \\
\hline
\end{tabular}
\end{center}
\caption{Results for the fit~(\ref{eq:critical_scaling_3}) 
for several values of $L_{\rm min}$. Here $\beta = 0.311$.
}
\label{results-fit-critical-scaling-2}
\end{table}  
They are now in good agreement with the theoretical 
prediction $\gamma/\nu = 2$. 
The same analysis has been performed for the correlation length.
Using
\be
\xi(\beta,L) = 
Z_\xi L^{\rho} \left(1 + X_\xi L^{2}\right),
\ee
for $L_{\min} = 20$ we obtain $\rho = 1.000(94)$, $Z_\xi =
0.1524(37)$, $X_\xi = 0.00(96)\cdot 10^{-4}$ with
$\chi^2/(\text{DOF})=1.53$, $\text{DOF}=2$. In this case, the addition of the 
correction does not change the result.

As a further check we have repeated the fit~(\ref{eq:critical_scaling_3}) by 
using the data at $\beta=0.3105$. 
The results are reported in Table~\ref{results-fit-critical-scaling-3}.
\begin{table}
\begin{center}
\begin{tabular}{|l|l|l|l|l|l|}
\hline
$L_{\rm min}$ & $\gamma/\nu$ & $Z_\chi$ &$X_\chi$ & $\chi^2$/DOF & DOF \\
\hline
14 & 2.061(50)  & 0.0734(82)  & $-$0.000169(55) & 10  & 4  \\
16 & 2.037(85)  & 0.078(17)    & $-$0.000147(90)   & 9.5  &3   \\
18 & 2.021(58) & 0.082(12)    & $-$0.000133(53)  & 13  & 2  \\
20 & 1.9(1.2)  &  0.11(37)   & 0.0000(11)  & 7.9  &  1 \\
\hline
\end{tabular}
\end{center}
\caption{Results for the fit~(\ref{eq:critical_scaling_3})
for several values of $L_{\rm min}$.  Here $\beta = 0.3105$.
}
\label{results-fit-critical-scaling-3}
\end{table}  
The estimates of $\gamma/\nu$ are in good agreement with the 
expected $\gamma/\nu= 2$. Also, the estimates of $X_\chi$ are 
in rough agreement with the prediction $X_\chi(\beta) \propto (\beta-\beta_c)$,
which implies $X_\chi(0.3105)\approx 4 X_\chi(0.3110)$. 
A combined fit of the data at $\beta=0.311$ and $\beta = 0.3105$ to
\be
\chi(\beta,L) = L^{\gamma/\nu} (1 + Y_\chi (\beta - \beta_c) L^2),
\ee
gives, by using our estimate \reff{betacfinale}, 
$\gamma/\nu = 2.032(80)$ (resp. $2.00(22)$) for $L_{\rm min} = 16$
(resp. 18),
with $\chi^2/\text{DOF}$ comparable to those of the previous fits. 
The agreement is clearly quite good.

In principle, one should also consider corrections to scaling proportional 
to $L^{-\omega}$,
where we expect $\omega\ge 2$ as a consequence of our 
findings $\theta\ge 1$. However, our data are not precise enough 
to allow for the determination of such a term.

\section{Conclusions}
\label{sec:conclusions}

In this paper we have performed a thorough check of the theoretical predictions
for the DLG. Assuming $\Delta = 2$, we checked that the transverse 
susceptibility $\chi$ and the transverse correlation length $\xi_{13}$
have the correct behavior both in infinite volume for $\beta\to\beta_c$
and at the critical point for $L\to\infty$. In other words, 
we find very good agreement with the field-theoretical predictions 
\reff{sec4:infvol} and \reff{sec4:FSS}. Concerning the exponent $\beta$, 
the analysis of the 
magnetization at the critical point gives,
although with limited precision,
$\beta\approx 1/2$, in agreement
with the field-theoretical results. It is important to notice that 
in all these analyses we have not found any evidence for the presence
of logarithmic corrections. As it has observed in previous studies
\cite{Leung92}, if they are really there, they are quite small.

Our result for the Binder parameter $g$\ does not agree with that
of Ref. \cite{Leung92} where it was found
$g\not=0$ at criticality, but confirms the 
results of Wang \cite{Wang96} that could not find a satisfactory
collapse for the Binder parameter. Our result $g=0$ is compatible
with the idea that in the scaling limit transverse correlations
(both in infinite volume and in the finite-size scaling regime)
are Gaussian, so that $g=0$ at criticality. Such a conclusion is also
supported by the analysis of the distribution function of the finite-volume
magnetization that is perfectly compatible with a purely
Gaussian behavior.
Note that this is
not in contrast with the fact that $g\not=0$ in the low-temperature
phase. Indeed, it is possible that $g$ behaves as the magnetization,
i.e. $g \sim (T_c - T)^{\zeta}$ for $T\to T_c^-$, 
with $\zeta/\nu = q \approx 0.5$.
The observed exponent $q$ is however difficult to explain. 
We mention that it is
also possible that $g$ decreases as a power of $\log L$ because of the
marginal operator, but that, in our range of values of $L$,
the complicated logarithmic dependence is mimicked by a single power.      
Note that, if $g_\infty(\beta_c) = 0$, the Binder parameter 
cannot be used to compute $\beta_c$: The crossing method does not work.

There is still a question that we don't fully understand. 
While the critical behavior of renormalized observables is fully 
in agreement with the predictions of the corresponding effective 
field theory, the FSS functions that we obtain for the microscopic 
model from our MC simulations are related to those 
of the coarse-grained effective field theory by the nontrivial 
scaling functions $\tilde Z$\ and $\tilde\tau$\ 
(see Eqs.~\reff{tildezeta}\ and \reff{tildetau}). 
In particular, the relation between the reduced temperature $t$\ 
of the microscopic model and the ``temperature'' $\tau$\ 
of the coarse-grained one seems to be not analytically predictable. 
Is this a consequence of the neglected marginal operator? 
An analysis of the three-dimensional case (where the operator is irrelevant), 
already in progress, should help us to gain a deeper understanding.


\begin{thebibliography}{99}
\bibitem{foot1a}
The \dlg\ is soluble for infinite drive in the limit in which the ratio of 
jump rates parallel and perpendicular to the field direction becomes infinite 
\cite{vB-S-84}.
\bibitem{foot4}
In principle, it is enough to consider periodic boundary conditions in the
field direction. The boundary conditions in the transverse directions
are largely irrelevant for the problems discussed here.
\bibitem{foot2a}
A field theory for the DLG was also derived in Ref.~\cite{GawKupi86},
starting from the standard Model-B dynamics.
The external drive partially breaks the supersymmetry (SUSY) of Model B, 
giving rise to a crossover towards a new fixed point in $d=5$, 
with a residual symmetry ($\frac{1}{2}$SUSY of Ref.~\cite{GawKupi86}).
\bibitem{foot5}
We must note that the susceptibility  defined by using the
  linear response theory does not coincide in general non-equilibrium system
  with that defined in terms of the Fourier transform of the two-point
  correlation function.
\bibitem{foot6}
Not any basis in the space spanned by the functions $\{ z^k \}$ is suitable to
perform the fit. As discussed in Ref. \cite{Mana-etal-97},
care must be taken to choose functions which are
effectively orthogonal with respect to the scalar product induced by the
$\chi^2$, otherwise the numerical solution
of the minimization problem can be unstable.
We used the functions
$
f_k(z) = z^2 P^{(6,2)}_{k-1}(2 z/z_{\text{max}}-1) ,
$
where $P^{(a,b)}_n$ are Jacobi polynomials which are an orthonormal system in
$L^2([0,1],x^a (1-x)^b)$.The constant $z_{\text{max}}$ must be such that
$z_{\text{max}} > z^a_j$ for all $a,j$. We chose $z_{\text{max}}=0.16$.
The choice of the parameters $a,b$ is
not crucial and different choices would perform analogously.
\end{thebibliography}

\newpage

\end{document}